\documentclass[english,aps,preprintnumbers,amsmath,amssymb,nofootinbib,twocolumn,10pt,prd]{revtex4-1} 

\usepackage[utf8]{inputenc}
\usepackage[T1]{fontenc}

\usepackage{ifthen}
\usepackage{forloop}
\usepackage{babel}
\usepackage{graphicx}
\usepackage{psfrag}
\usepackage{color}
\usepackage{dsfont}
\usepackage{mathrsfs}

\usepackage{slashed}

\usepackage{mathbbol}

\usepackage{ulem}

\usepackage{color}

\renewcommand{\Re}{\operatorname{Re}}

\newcommand{\metric}{\ensuremath{g}}
\newcommand{\vierbein}{\ensuremath{e}}
\newcommand{\christoffel}[1]{\text{\footnotesize$\left\{ \begin{matrix} #1 \end{matrix} \right\}$}}
\newcommand{\Obs}{\ensuremath{O}}
\newcounter{pointcounter}
\newcommand{\point}[1][\empty]{
  \ifthenelse
    {\equal{#1}{\empty}}
    {\ensuremath{\, \phantom{\cdot} \,}}
    {\setcounter{pointcounter}{1} \forloop{pointcounter}{0}{\value{pointcounter} < #1}{\ensuremath{\, \phantom{\cdot} \,}}}
}

\newcommand{\Eqref}[1]{Eq.~\eqref{#1}}
\newcommand{\ttm}[1]{\ensuremath{\text{\tiny{$#1$}}}}
\newcommand{\euler}{\mathrm{e}}
\newcommand{\cplx}{\mri}

\newcommand{\tr}{\operatorname{tr}}

\newcommand{\abs}[1]{\ensuremath{\left\vert#1\right\vert}}
\newcommand{\regint}[1]{\int \!\!\! {}_{{}_{{}_{{}_{{}_\text{\small{\ensuremath{#1}}}}}}}}
\newcommand{\mcA}{\ensuremath{\mathcal{A}}}

\newcommand{\mcD}{\ensuremath{\mathcal{D}}}

\newcommand{\mcF}{\ensuremath{\mathcal{F}}}
\newcommand{\mcG}{\ensuremath{\mathcal{G}}}

\newcommand{\mcL}{\ensuremath{\mathcal{L}}}
\newcommand{\mcM}{\ensuremath{\mathcal{M}}}

\newcommand{\mcO}{\ensuremath{\mathcal{O}}}

\newcommand{\mcS}{\ensuremath{\mathcal{S}}}

\newcommand{\mcZ}{\ensuremath{\mathcal{Z}}}

\newcommand{\mrd}{\ensuremath{\mathrm{d}}}

\newcommand{\mri}{\ensuremath{\mathrm{i}}}
\newcommand{\mrI}{\ensuremath{\mathrm{I}}}

\newcommand{\mrT}{\ensuremath{\mathrm{T}}}

\newcommand{\mfrg}{\ensuremath{\mathfrak{g}}}

\newcommand{\mfrL}{\ensuremath{\mathfrak{L}}}

\newcommand{\C}{\ensuremath{\mathds{C}}}
\newcommand{\N}{\ensuremath{\mathds{N}}}
\newcommand{\R}{\ensuremath{\mathds{R}}}

\newcommand{\rmi}{\ensuremath{(\mathrm{i})}}
\newcommand{\rmii}{\ensuremath{(\mathrm{ii})}}
\newcommand{\rmiii}{\ensuremath{(\mathrm{iii})}}
\newcommand{\rmiv}{\ensuremath{(\mathrm{iv})}}
\newcommand{\rmv}{\ensuremath{(\mathrm{v})}}

\makeatother

\begin{document}

\title{Fermions in gravity with local spin-base invariance 
}
\author{Holger Gies and Stefan Lippoldt}
\affiliation{Theoretisch-Physikalisches Institut, Friedrich-Schiller-Universit\"at Jena, 
Max-Wien-Platz 1, D-07743 Jena, Germany}                                        

\begin{abstract}

We study a formulation of Dirac fermions in curved spacetime that
respects general coordinate invariance as well as invariance under
local spin-base transformations. The natural variables for this
formulation are spacetime-dependent Dirac matrices subject to the
Clifford-algebra constraint. In particular, a coframe, i.e. vierbein
field is not required. The corresponding affine spin connection
consists of a canonical part that is completely fixed in terms of the
Dirac matrices and a free part that can be interpreted as spin
torsion. A general variation of the Dirac matrices naturally induces a
spinorial Lie derivative which coincides with the known Kosmann-Lie
derivative in the absence of torsion. Using this formulation
for building a field theory of quantized gravity and matter fields, we
show that it suffices to quantize the metric and the matter
fields. This observation is of particular relevance for field theory
approaches to quantum gravity, as it can serve for a purely
metric-based quantization scheme for gravity even in the presence of
fermions.

\end{abstract}

\maketitle

\section{Introduction}\label{sec:intro}

Building a field theory of quantized gravity requires to specify the
fundamental degrees of freedom to be quantized. Unfortunately, the guidance
from the corresponding classical theory, general relativity, is not
particularly strong, as the variational principle applied to substantially
different degrees of freedom can lead to the same equations of
motion. Examples are given by (i) the conventional Einstein-Hilbert action in
terms of metric degrees of freedom $g_{\mu\nu}$, (ii) or in terms of a
vierbein $\vierbein_{\mu}^{\point a}$, (iii) or the first-order Hilbert
Palatini action which in addition to the vierbein also depends on the spin
connection $\omega_{\mu \point b}^{\point a}$ (Einstein-Cartan theory). Many
further variants along this line are known \cite{Ashtekar:2004eh}, all of
which (in the absence of torsion or other deformations) have in common that
they yield Einstein's equation on the classical level.

By contrast, if these various classically equivalent theories are quantized
(by some appropriate method), the quantum versions should be expected to
generically differ. This can be seen from the fact that, e.g., the relation
between the metric and the vierbein, 
\begin{equation}
 \metric_{\mu \nu} = \vierbein_{\mu}^{\point a} \vierbein_{\nu}^{\point b}
 \eta_{a b},
\label{eq:gande}
\end{equation}
implies that an appropriate measure for a functional integral over metrics
$\mathcal{D} g$ is expected to differ from that over vierbeins $\mathcal{D} e$
by a nontrivial Jacobian. The resulting differences have explicitly been
worked out recently in the asymptotic safety approach to quantum gravity
\cite{Weinberg:1980gg,Reuter:1996cp}, but should be expected to occur in any other field
theory attempt at quantizing gravity as well. For instance, the RG flow of
metric-based quantum gravity \cite{Reuter:1996cp} has been shown to differ from
that of its vierbein-based counterpart \cite{Harst:2012ni,Dona:2012am} at least
quantitatively. Qualitatively, new aspects arise from the Faddeev-Popov ghosts
associated with local Lorentz invariance in the vierbein formulation
\cite{Harst:2012ni} -- a symmetry that is not present in the metric-based
formulation. In the same spirit, quantizing Einstein-Cartan theory or its
chiral variants leads to yet further sets of RG flows
\cite{Daum:2010qt,Harst:2013} even in the absence of any torsion. 

As only one of these different quantum theories can be realized in Nature,
criteria beyond pure mathematical consistency are required to distinguish
between the different theories. As fermions occur in our universe, the use of
a vierbein-based formalism seems mandatory, consequently giving preference to
versions of quantum gravity where the corresponding fields are considered as
fundamental or where at least vierbeins are formed prior to the metric in
terms of even more fundamental degrees of freedom, see
e.g., \cite{Hebecker:2003iw,Diakonov:2011im}.

In the present work, we critically reexamine the seeming necessity of
vierbein-based formulations in the presence of fermions on curved
space.  For this, we consider a more general formulation of fermions
in gravity, where in addition to general coordinate invariance the
symmetry under local spin-base transformations remains fully preserved
\cite{Finster:1997gn,Weldon:2000fr}. With respect to our original
motivation, it turns out that a purely metric-based quantization
scheme appears much more natural, as the local spin-base
fluctuations can be shown to represent a trivial factor of the
measure. We emphasize that this observation does not invalidate a
quantization of gravity in terms of vierbeins or other underlying
degrees of freedom. Rather, the existence of fermions in the universe
does not provide an argument for ruling out metric-based quantization
schemes of gravity. A similar conclusion has been drawn for the case
that the observed fermions finally turn out to be K\"{a}hler fermions
\cite{Dona:2012am}.

On our way to this central result, we will re-derive and generalize
the spin-base invariant formalism for fermions in curved space,
following the work of Finster and Weldon
\cite{Finster:1997gn,Weldon:2000fr}. In particular, we derive all
details of the formalism as well as new results from very few
underlying assumptions in a self-contained way. Among the new results,
we show how the concept of spin torsion arises in this formalism and
we discover a new simple relation between the general variation of
Dirac matrices and the Kosmann-Lie derivative for spinors.

In order to contrast the spin-base invariant formalism discussed below
with the standard vierbein formulation, let us briefly recall the
elements of the standard construction for describing fermions in
curved spacetime \cite{Weyl:1929,Fock:1929,DeWitt:1965jb,Buchbinder:1992rb}: once a
suitable vierbein, satisfying \Eqref{eq:gande} is introduced, the spin
connection $\omega_{\mu \point b}^{\point a}$ required to define
fermionic dynamics is derived from the vierbein postulate as an
algebraic equation
\begin{equation}\label{eq:vierbein_postulate}
 0 = \partial_{\mu} \vierbein_{\nu}^{\point a} - \Gamma_{\mu \nu}^{\kappa} \vierbein_{\kappa}^{\point a} + \omega_{\mu \point b}^{\point a} \vierbein_{\nu}^{\point b} \text{,}
\end{equation}
where $\Gamma_{\mu \nu}^{\kappa}$ is the -- not necessarily symmetric
-- affine spacetime connection
\cite{DeWitt:1965jb,Buchbinder:1992rb,Watanabe:2004nt}.  The Dirac
matrices ${\gamma_\ttm{(\vierbein)}}_{\mu}$ within this vierbein
formalism are given by
\begin{equation}
 {\gamma_\ttm{(\vierbein)}}_{\mu} = \vierbein_{\mu}^{\point a} {\gamma_\ttm{(\mathrm{f})}}_{a} \text{,}
\label{eq:gammae}
\end{equation}
where the ${\gamma_\ttm{(\mathrm{f})}}_{a}$ are fixed constant Dirac matrices satisfying the Clifford algebra for Minkowski space
\begin{equation}
 \{ {\gamma_\ttm{(\mathrm{f})}}_{a} , {\gamma_\ttm{(\mathrm{f})}}_{b} \} = 2 \eta_{a b} \mrI \text{,}
\end{equation}
where $\mrI$ is the unit matrix. In this way, the
${\gamma_\ttm{(\vierbein)}}_{\mu}$ are automatically compatible with
the Clifford algebra
\begin{equation}
 \{ {\gamma_\ttm{(\vierbein)}}_{\mu} , {\gamma_\ttm{(\vierbein)}}_{\nu} \} = 2 \metric_{\mu \nu} \mrI \text{.}
\end{equation}
The covariant derivative for spinors $\psi$ then reads
\begin{equation}
 {\nabla_\ttm{(\vierbein)}}_{\mu} \psi = \partial_{\mu} \psi + \frac{1}{8} \omega_{\mu}^{\point a b} [{\gamma_\ttm{(\mathrm{f})}}_{a}, {\gamma_\ttm{(\mathrm{f})}}_{b}] \psi \text{.}
\end{equation}
For explicit calculations the Dirac operator
$\slashed{\nabla}_\ttm{(\vierbein)} = {\gamma_\ttm{(\vierbein)}}^{\mu}
{\nabla_\ttm{(\vierbein)}}_{\mu}$ is often needed. In practical
calculations, it can be more convenient to have the Dirac operator in
a more adjusted basis concerning the actual choice of the Dirac
matrices \cite{Finster:1998ws,Casals:2012es}.

While this standard vierbein formalism is perfectly sufficient for a
description of fermions in curved spacetime, several properties give
rise to criticism at least from a conceptual (or aesthetic) viewpoint:
the relevant objects for the physical system are the generally
spacetime-dependent Dirac matrices $\gamma_\mu$ which have to satisfy
the Clifford algebra $\{ \gamma_\mu,\gamma_\nu\} =2 \metric_{\mu\nu}
\mrI$. For a given metric, more solutions than only those
parameterizable by a vierbein exist for the Dirac matrices
\cite{Weldon:2000fr}. This already indicates that the vierbein
construction should be regarded as a special choice. On the other
hand, it seems at odds with the principles of general relativity that
a special inertial coframe $\vierbein_{\mu}^{\point a}$ has to be
introduced in order to describe the fermions.

In addition, this choice introduces another symmetry, ``physically''
corresponding to the Lorentz symmetry of the tangential space related
to the roman indices $a,b, \dots$ (e.g., a local O(4) symmetry in a
Euclidean formulation, which can be generalized to a GL($4,\R$) 
symmetry \cite{Floreanini:1989hq}). From the viewpoint of the Dirac
matrices $\gamma_\mu$, this symmetry seems artificial. By contrast,
the relevant nontrivial symmetry of the Clifford algebra is the local
spin-base symmetry SL(4,$\C$) which is not fully reflected by the
standard vierbein construction.

The spin-base invariant formalism \cite{Finster:1997gn,Weldon:2000fr}
used and further developed in the present work does not require a
coframe or vierbein construction. Still, in the absence of torsion it
is completely compatible with the vierbein formalism in the sense that
a vierbein construction can always be recovered as a special
case. Important differences however arise in the presence of torsion,
as discussed in Sect.~\ref{sec:Consequ}. The spin-base invariant
formalism supports degrees of freedom within the affine spin
connection, which can be interpreted as a spin torsion. This is in
direct analogy to the affine spacetime connection which in general
consists of a canonical part in terms of the Levi-Civita connection and a
free part connected with spacetime torsion.

Following the principle of general covariance together with spin-base
invariance, a field strength corresponding to a spin curvature can be
constructed. The simplest action linear in this field strength defines
a classical dynamical theory. The resulting equations of motion imply
that the spin torsion vanishes in the absence of any sources. The
metric-part of these equations of motion correspond to general
relativity as expected. 

This paper is organized as follows: in Sect. \ref{sec:Axioms}, we
specify all prerequisites and assumptions for constructing the
spin-base invariant formalism. Section \ref{sec:Consequ} is devoted to
the analysis of the affine spin connection and the spin metric, which
defines the relation between spinors and Dirac conjugated spinors. In
Sect. \ref{sec:lie} a spinorial Lie derivative is constructed within
the present framework which turns out to coincide with the Kosmann-Lie
derivative known in the literature. The inclusion of an additional
gauge symmetry is worked out in Sect. \ref{sec:gauge_field}.  The
field strength for spinors and the corresponding action linear in the
field strength is derived in Sect. \ref{sec:spin_curv}.  We generalize
our results, formulated for irreducible representations of the Dirac
algebra, to reducible cases in Sect. \ref{sec:red_rep}. The
implications of the spin-base invariant formalism for a possible
quantized version of gravity and quantized matter is discussed in
Sect. \ref{sec:pathintegral} on the level of a path integral
approach. As a first hands-on application of the spin-base invariant
formalism, we determine the response of several elements of the
formalism (Dirac matrices, spin connection, etc.) under a variation of
the metric in Sect. \ref{sec:variations}. These results form
elementary technical building blocks for generic quantum field theory
computations. Conclusions are drawn in Sect. \ref{sec:conc}. The
uniqueness (up to a sign) of the spin metric is proven in
App. \ref{App:spin_metric}. In App. \ref{App:formulae} we list several
useful identities of the formalism for the simpler case of vanishing
torsion, serving as a toolbox for a straightforward application of the
formalism.

\section{Basic requirements for spin-base invariance}
\label{sec:Axioms}

We aim at a generally covariant and spin-base invariant description of
fermions without recourse to a vierbein construction. For this, only a
few basic assumptions have to be made. We stress that these
requirements are completely compatible with the vierbein formalism for
torsionfree spacetimes.

First we fix the relation between the metric of a (pseudo-)Riemannian
spacetime and the Dirac matrices $\gamma^{\mu}$ by demanding the
Clifford algebra to hold locally,
\begin{align}
 \{ \gamma^{\mu} , \gamma^{\nu} \} = 2 \metric^{\mu \nu} \mrI , \quad \gamma^{\mu} \in \C^{d_{\gamma} \times d_{\gamma}}.
\label{def:Clifford}
\end{align}
Here, $d_\gamma$ denotes the dimension of the Dirac matrices in the
\textit{irreducible} representation of the Clifford algebra, i.e., $d_\gamma=2^{\lfloor d/2 \rfloor}$.

The Clifford algebra supports an $\text{SL}(d_{\gamma},\C)$ symmetry.%
\footnote{In fact the Clifford algebra is invariant under a
  $\text{GL}(d_{\gamma},\C)$ symmetry which locally factorizes into
  $\text{SL}(d_{\gamma},\C) \times \text{U}(1) \times \R_+$.  Here we
  first concentrate on the $\text{SL}(d_{\gamma},\C)$ component, as the
  $\text{U}(1) \times \R_+$ part does act trivially on the Dirac matrices.
  The inclusion of additional symmetry groups such as the U(1) factor is discussed in
  Sect. \ref{sec:gauge_field}.}
We require this invariance under \textit{spin-base transformations}
with $\mcS \in \text{SL}(d_{\gamma},\C)$ to hold locally for general
action functionals involving the Dirac matrices and Dirac fermions
$\psi$ and their conjugate $\bar{\psi}$ obeying the transformation
rules,
\begin{align}
\begin{aligned}
 \rmi{}&\\
  \rmii{}&\\
  \rmiii{}&
\end{aligned} \quad
\begin{aligned}
 \gamma^{\mu} {}&\rightarrow \mcS \gamma^{\mu} \mcS^{-1} \text{,}\\
 \psi {}&\rightarrow \mcS \psi \text{,}\\
 \bar{\psi} {}&\rightarrow \bar{\psi} \mcS^{-1} \text{.}
\end{aligned}
\end{align}
Dirac conjugation of a spinor $\psi$ involves hermitean conjugation
and a \textit{spin metric} $h$,
\begin{align}\label{eq:def_spin_metric}
 {}&\bar{\psi} = \psi^{\dagger} h \text{,}
\end{align}
which is assumed to carry no scale,
\begin{align}
 {}&\abs{\det h} = 1 \text{.} \label{eq:det_spin_metric}
\end{align}
Local spin-base invariance requires to introduce a covariant
derivative $\nabla_\mu$ with the following standard properties,
\begin{align}\label{eq:axiom_cov_deriv}
 \begin{aligned}
  \rmi{}&\\
  \rmii{}&\\
  \rmiii{}&
 \end{aligned} \,
 \begin{aligned}
  &\text{linearity:}\\
  &\text{product rule:}\\
  &\text{covariance:}
 \end{aligned} \,\,
 \begin{aligned}
  &\nabla_{\mu}( \psi_{1} + \psi_{2} ) = \nabla_{\mu} \psi_{1} + \nabla_{\mu} \psi_{2} \text{,}\\
  &\nabla_{\mu}(\psi \bar{\psi}) = (\nabla_{\mu} \psi) \bar{\psi} + \psi (\nabla_{\mu} \bar{\psi}) \text{,}\\
  &\nabla_{\mu} \bar{\psi} = \overline{\nabla_{\mu} \psi}, \quad \nabla_{\mu} \psi^{\dagger} = (\nabla_{\mu} \psi)^{\dagger}\text{.}
 \end{aligned}
\end{align}
Up to this point, local spin-base invariance is reminiscent to gauge invariance,
however, with a non-compact gauge group. Now we make contact with general
covariance by additionally demanding that $\nabla_{\mu}$ has to
coincide with the ordinary spacetime covariant derivative $D_{\mu}$,
if it acts on an object that is a scalar under spin-base
transformations. A particularly important example is given by
\begin{align}
 \nabla_{\mu} (\bar{\psi} \gamma^{\nu} \psi) = D_{\mu} (\bar{\psi} \gamma^{\nu} \psi) = \partial_{\mu} (\bar{\psi} \gamma^{\nu} \psi) + \Gamma^{\nu}_{\mu \kappa} (\bar{\psi} \gamma^{\kappa} \psi) \text{,}
\label{def:gencov}
\end{align}
where $\Gamma^{\nu}_{\mu \kappa}$ denotes the \textit{metric compatible} affine spacetime connection
\begin{align}
 \Gamma^{\nu}_{\mu \kappa} = \christoffel{\nu \\ \mu \kappa} + K^{\nu}_{\point \mu \kappa}.
\label{def:spacetimeconnection}
\end{align}
Here $\christoffel{\nu \\ \mu \kappa}$ is the Levi-Civita connection
\begin{align}
 \christoffel{\nu \\ \mu \kappa} = \frac{1}{2} g^{\nu \lambda} \left( \partial_{\mu} g_{\lambda \kappa} + \partial_{\kappa} g_{\lambda \mu} - \partial_{\lambda} g_{\mu \kappa} \right),
\end{align}
and $K^{\nu}_{\point \mu \kappa}$ is the contorsion tensor. The contorsion $K^{\nu}_{\point \mu \kappa}$ and the torsion $C^{\nu}_{\point \mu \kappa}$ are related via
\begin{align}
 C^{\nu}_{\point \mu \kappa} ={}& 2 K^{\nu}_{\point[1] [\mu \kappa]} \equiv K^{\nu}_{\point \mu \kappa} - K^{\nu}_{\point \kappa \mu} \text{,}\\
 K^{\nu}_{\point \mu \kappa} ={}& \frac{1}{2} (C^{\nu}_{\point \mu \kappa} + C^{\point \nu}_{\kappa \point \mu} - C^{\point \point \nu}_{\mu \kappa}) \equiv - K_{\kappa \mu}^{\point \point \nu}\text{,}
\end{align}
where indices in square brackets $[\ldots]$ are completely antisymmetrized.

The property $\rmiii$ of \Eqref{eq:axiom_cov_deriv} can be interpreted
as the spinorial analog to metric compatibility of general relativity
with $\bar{\psi}$ corresponding to a covariant spin vector and $\psi$
to a contravariant spin vector under spin-base transformations. From
this viewpoint, the hermitean conjugate spinor $\psi^{\dagger}$ should
be considered as (the hermitean conjugate of) a contravariant spin
vector in contrast to the covariant $\bar{\psi}$. For instance,
$\psi^\dagger$ transforms with $\mcS^\dagger$, whereas $\bar\psi$
transforms with $\mcS^{-1}$. Therefore, we need the additional
\textit{definition} $\nabla_{\mu} \psi^{\dagger} = (\nabla_{\mu}
\psi)^{\dagger}$ in $\rmiii$ of \Eqref{eq:axiom_cov_deriv} which
reduces to an identity for the ordinary partial derivative in flat
space.%
\footnote{In fact this \textit{definition} is not mandatory. If we
  dropped $\nabla_{\mu} \psi^{\dagger} = (\nabla_{\mu}
  \psi)^{\dagger}$, there would be no unique definition of the
  spin covariant derivative for $\psi^{\dagger}$ and $h$. These
  derivatives are, however, not necessary for calculational or
  conceptual concerns. With hindsight, only the second
  equality in \Eqref{eq:cov_deriv_spin_metric} given below,
  $\partial_{\mu} h - h \Gamma_{\mu} - \Gamma_{\mu}^{\dagger} h = 0$,
  is needed which can be inferred from the property $\nabla_{\mu}
    \bar{\psi} = \overline{\nabla_{\mu} \psi}$ alone.  }

Finally,  we require the action of a dynamical theory to be real, especially
\begin{align}\label{eq:real_action}
\begin{aligned}
 \rmi{}& \quad (\bar{\psi} \psi)^{\ast} = \bar{\psi} \psi \text{,}\\
 \rmii{}& \quad \regint{x} (\bar{\psi} \slashed{\nabla} \psi)^{\ast} = \regint{x} \bar{\psi} \slashed{\nabla} \psi \text{,}
\end{aligned}
\end{align}
where $\regint{x}$ is a shorthand for $\int \mrd^d x \sqrt{- g}$ with
$g = \det g_{\mu \nu}$. Equations \eqref{eq:real_action} also fix some of our
conventions, i.e., in other conventions the reality conditions could
look differently.

Based on these elementary requirements, the next section is devoted to
the analysis of the spin connection that fully implements
invariance under spin-base transformation. Apart from obvious
conceptual advantages, this invariance might be of use in practical
computations to choose a convenient set of Dirac matrices for a
simplified construction of classical solutions
\cite{Finster:1998ws,Casals:2012es}. For vanishing torsion, this spin
connection can be made to agree with the standard spin connection
constructed from the vierbeins $\vierbein^{\mu}_{\point a}$ if the
Dirac matrices are spin-base transformed to those of the vierbein
formalism.

\section{Properties of the affine spin connection}
\label{sec:Consequ}

For the following analysis, we work in $d=4$ spacetime
dimensions, where $d_\gamma=4$ holds for the dimension of the
irreducible representation of the Clifford algebra. Generalizations to
$d=2$ and $d=3$ can also be worked out straightforwardly
\cite{Lippoldt:2012}. Whenever suitable, we give the formulas for
general $d_\gamma$ to emphasize the generality of parts of the
construction.

A cornerstone of the present construction is the Weldon theorem
\cite{Weldon:2000fr}. It states that an infinitesimal variation of the
Dirac matrices which preserves the Clifford algebra can be decomposed
into an infinitesimal variation of the inverse metric $\delta g^{\mu \nu}$ and
an infinitesimal SL$(4,\C)_\gamma$ transformation $\delta \mcS_{\gamma}$:
\begin{align}
 \delta \gamma^{\mu} = \frac{1}{2} (\delta \metric^{\mu \nu}) \gamma_{\nu} + [\delta \mcS_{\gamma} , \gamma^{\mu}] \text{.} \label{eq:variation_gamma}
\end{align}
Here the notation distinguishes between SL($4,\C$) spin-base
transformations $\mcS$ that simultaneously act on fermions as
discussed above and constitute an invariance of the theory, and
an (infinitesimal) SL$(4,\C)_\gamma$ transformation
$\delta\mcS_\gamma$ that may only act on the Dirac
matrices.%
\footnote{In a slight abuse of language, we may call $\delta
  \mcS_\gamma$ a ``spin-base fluctuation''. Since it applies only to
  the Dirac matrices, it represents a physically relevant fluctuation
  of the spin basis for the Dirac matrices relative to that of the
  fermions. This is different from an infinitesimal version of the spin-base
  transformation $ \mcS\in$ SL($4,\C$) which are an invariance of
  the theory by definition.}
This theorem can
straightforwardly be proved by using that every matrix in
$\C^{4 \times 4}$ can uniquely be locally spanned by
the explicit basis of the Clifford algebra $\mrI$, $\gamma_{\ast}$,
$\gamma_{\alpha}$, $\gamma_{\ast} \gamma_{\alpha}$ and
$[\gamma_{\alpha} , \gamma_{\beta}]$. Here,
\begin{align}
 \gamma_{\ast} ={}& - \frac{\cplx}{4!} \tilde{\varepsilon}_{\mu_{1} \ldots \mu_{4}} \gamma^{\mu_{1}} \ldots \gamma^{\mu_{4}}, \quad \tilde{\varepsilon}_{\mu_{1} \ldots \mu_{4}} = \sqrt{-g} \varepsilon_{\mu_{1} \ldots \mu_{4}} \text{}
\end{align}
is the generalized (generally spacetime-dependent) analog of $\gamma_5
= - \cplx {\gamma_\ttm{(\mathrm{f})}}^{0}
{\gamma_\ttm{(\mathrm{f})}}^{1} {\gamma_\ttm{(\mathrm{f})}}^{2}
{\gamma_\ttm{(\mathrm{f})}}^{3}$ in flat space. We have used the
Levi-Civita symbol $\varepsilon_{\mu_{1} \ldots \mu_{4}}$ with
$\varepsilon_{0 1 2 3} = 1$ to define the Levi-Civita tensor
$\tilde{\varepsilon}_{\mu_{1} \ldots \mu_{4}}$. Essential properties
of $\gamma_\ast$ are
\begin{align}
\begin{aligned}
 \rmi \quad{}& \{ \gamma^{\mu} , \gamma_{\ast} \} = 0 \text{,}\\
 \rmii \quad{}& \tr \gamma_{\ast} = 0 \text{.}
\end{aligned}
\end{align}
Now, let us assume that a spacetime is specified in terms of a set of
Dirac matrices $\gamma_\mu$, also defining the metric through the
Clifford algebra \Eqref{def:Clifford}, and in terms of contorsion
$K^\nu_{\point\mu\kappa}$. Already from the Dirac matrices, we can
determine a useful auxiliary Dirac-valued matrix $\hat{\Gamma}_{\mu}$
as a spacetime vector valued function of the $\gamma^{\mu}$. It is
defined by
\begin{align}\label{eq:def_hatGamma}
 {D_\ttm{(\mathrm{LC})}}_{\mu} \gamma^{\nu} = \partial_{\mu} \gamma^{\nu} \! + \! \christoffel{\nu \\ \mu \kappa} \gamma^{\kappa} ={}& \! - [ \hat{\Gamma}_{\mu} , \gamma^{\nu} ]\text{,} \,\, \tr \! \hat{\Gamma}_{\mu} = \! 0,
\end{align}
where ${D_\ttm{(\mathrm{LC})}}_{\mu}$ is the spacetime covariant
derivative including the Levi-Civita connection, but disregarding any
torsion. Equations \eqref{eq:def_hatGamma} can be resolved explicitly
in terms of the local Clifford basis:
\begin{align}
\begin{aligned}
 \rmi \quad{}& \hat{\Gamma}_{\mu} = p_{\mu} \gamma_{\ast} + v_{\mu}^{\point \alpha} \gamma_{\alpha} + a_{\mu}^{\point \alpha} \gamma_{\ast} \gamma_{\alpha} + t_{\mu}^{\point \alpha \beta} [ \gamma_{\alpha} , \gamma_{\beta} ] \text{,}\\
 \rmii \quad{}& p_{\mu} = \frac{1}{32} \tr(\gamma_{\ast} \gamma_{\alpha} \partial_{\mu} \gamma^{\alpha}) \text{,}\\
 \rmiii \quad{}& v_{\mu}^{\point \alpha} = \frac{1}{48} \tr( [ \gamma^{\alpha} , \gamma_{\beta} ] \partial_{\mu} \gamma^{\beta}) \text{,}\\
 \rmiv \quad{}& a_{\mu}^{\point \alpha} = \frac{1}{8} \tr( \gamma_{\ast} \partial_{\mu} \gamma^{\alpha} ) \text{,}\\
 \rmv \quad{}& t_{\mu \alpha}^{\point[2] \beta} = - \frac{1}{32} \tr( \gamma_{\alpha} \partial_{\mu} \gamma^{\beta} ) - \frac{1}{8} \christoffel{\beta \\ \mu \alpha} \equiv - t_{\mu \point \alpha}^{\point \beta} \text{,}
\end{aligned}\label{eq:spintorsionconstraints}
\end{align}
where all tensorial coefficients are functions of the Dirac matrices. 

Next, we turn to the construction of the covariant derivative. From  $\rmii$ of
\Eqref{eq:axiom_cov_deriv} and $\nabla_{\mu} (\bar{\psi} \psi) = \partial_{\mu} (\bar{\psi} \psi)$
analogous to \Eqref{def:gencov}, we observe, that the covariant derivative can be written as 
\begin{align}
 \nabla_{\mu} \psi ={}& \partial_{\mu} \psi + \Gamma_{\mu} \psi \text{,}\\
 \nabla_{\mu} \bar{\psi} ={}& \partial_{\mu} \bar{\psi} - \bar{\psi} \Gamma_{\mu} \text{.}
\end{align}
Here we have introduced the affine spin connection $\Gamma_{\mu}$ to be analyzed,
which transforms as a vector under general coordinate transformations
and inhomogeneously under spin-base transformations, 
\begin{align}\label{eq:gauge_trafo_Gamma}
 \Gamma_{\mu} \rightarrow \mcS \Gamma_{\mu} \mcS^{-1} - ( \partial_{\mu} \mcS ) \mcS^{-1}.
\end{align}
From \Eqref{eq:def_hatGamma}, it is immediate that $\hat{\Gamma}_\mu$
has the same transformation properties as $\Gamma_\mu$ both under
general coordinate as well as spin-base transformations.  Because
$\Gamma_{\mu}$ is the connection for the spin-base transformations, it
is composed from the generators of the symmetry group SL$(d_{\gamma},
\C)$, the traceless matrices. Therefore we can set the trace of
$\Gamma_{\mu}$ to zero, $\tr \Gamma_{\mu}=0$. In
Sect. \ref{sec:gauge_field} below, we discuss generalizations
including a trace part.

From the property of general covariance \eqref{def:gencov}
together with the definition of Dirac conjugation in
\Eqref{eq:def_spin_metric}, we conclude that
\begin{align} 
 \nabla_{\mu} \gamma^{\nu} ={}& D_{\mu} \gamma^{\nu} \! + \! [ \Gamma_{\mu} , \gamma^{\nu} ] \equiv \! \left[ \Gamma_{\mu} \! - \! \hat{\Gamma}_{\mu} \! - \! \frac{1}{8} K_{\rho \mu \lambda} [\gamma^{\rho} , \gamma^{\lambda}] , \gamma^{\nu} \right] \! \text{,} \label{eq:cov_deriv_gamma}\\
 \nabla_{\mu} h ={}& \partial_{\mu} h - h \Gamma_{\mu} - \Gamma_{\mu}^{\dagger} h = 0 \text{,} \label{eq:cov_deriv_spin_metric}
\end{align}
where we have made use of the auxiliary matrix $\hat{\Gamma}_\mu$ 
defined in \Eqref{eq:def_hatGamma}.

The challenging task is to find the maximum number of constraints on
$\Gamma_{\mu}$ from the Dirac matrices and therefore from the metric
and the actual choice of the spin-base in order to identify its
physical content. For this, we first consider the spin metric and
notice with \Eqref{eq:def_spin_metric} that it transforms under
spin-base transformations as
\begin{align}
 h \rightarrow {\mcS^{\dagger}}^{-1} h \mcS^{-1} \text{.}
\end{align}
Eq. $\rmi$ of (\ref{eq:real_action}),
\begin{align}
 \psi^{\dagger} h \psi = \bar{\psi} \psi = (\bar{\psi} \psi)^{\ast} ={}& \psi^{\mrT} h^{\ast} \psi^{\ast} = \psi^{\dagger} ( - h^{\dagger}) \psi,
\end{align}
implies that the spin metric is antihermitean
\begin{align}
 h^{\dagger} ={}& -h \label{eq:spin_metric_antiherm} \text{.}
\end{align}
Let us now define the Dirac conjugate of a matrix $M$ as
\begin{align}
 \bar{M} = h^{-1} M^{\dagger} h
\end{align}
which implies
\begin{align}
 (\bar{\psi} M \psi)^{\ast} = \bar{\psi} \bar{M} \psi \text{.}
\end{align}
Using the standard relation
\begin{align}
 \quad \partial_{\mu} \sqrt{-\metric} = \sqrt{-\metric} \, \Gamma_{\mu \kappa}^{\kappa} \equiv \sqrt{-\metric} \, \christoffel{\kappa \\ \mu \kappa},
\end{align}
we can straightforwardly derive from Eq.~$\rmii$ of
(\ref{eq:real_action}) together with Eqs.~\eqref{eq:def_hatGamma} and
\eqref{eq:cov_deriv_spin_metric} that
\begin{align}
 \regint{x} \! \bar{\psi} \slashed{\nabla} \psi ={}& \! \regint{x} \! ( \bar{\psi} \slashed{\nabla} \psi )^{\ast} = \! \regint{x} \! \bar{\psi} \! \left( \! - \bar{\gamma}^{\mu} \nabla_{\mu} \! + \!\! \left[\bar{\Gamma}_{\mu} \! - \! \bar{\hat{\Gamma}}_{\mu} , \bar{\gamma}^{\mu}\right] \! \right) \! \psi.
\end{align}
As this has to hold for arbitrary fermion fields, we deduce
\begin{align}
 \bar{\gamma}^{\mu} = - \gamma^{\mu} \label{eq:gamma_dirac_conj}
\end{align}
and
\begin{align}
 0 = \left[ \Delta \Gamma_{\mu} , \gamma^{\mu} \right] \text{,} \quad \Delta \Gamma_{\mu} = \Gamma_{\mu} \! - \! \hat{\Gamma}_{\mu} \label{eq:def_Delta_Gamma} \text{.}
\end{align}
As $\hat{\Gamma}_\mu$ is fully determined in terms of the Dirac
matrices, \Eqref{eq:def_Delta_Gamma} represents a first constraint on
the components of the spin connection $\Gamma_\mu$. This constraint
can, of course, trivially be satisfied by identifying
$\Gamma_\mu\stackrel{!}{=}\hat{\Gamma}_{\mu}$ and setting the
difference to zero $\Delta \Gamma_\mu\stackrel{!}{=}0$. From the
present viewpoint, this is a perfectly legitimate choice, yielding one
particular explicit realization of the spin connection being fully
determined by the Dirac matrices. This choice has been advocated in
\cite{Weldon:2000fr}, where it has also been shown that this spin-base
invariant formalism contains the standard vierbein formalism as a
subset: for Dirac matrices following the vierbein construction
\Eqref{eq:gammae}, the coefficients $p_{\mu}, v_{\mu}^{\point \alpha},
a_{\mu}^{\point \alpha}$ all vanish, and the $t_{\mu}^{\point \alpha
  \beta}$ depend on 6 real parameters.

However, there is a priori no reason to single out this definition of
the spin connection in terms of $\hat{\Gamma}_\mu$. Therefore, we
investigate below the properties and possible further degrees of
freedom contained in a possibly nonzero $\Delta \Gamma_\mu$.

Before we do so, let us extract an important consequence of
\Eqref{eq:def_Delta_Gamma}: the covariant derivative of the Dirac
matrices given in \Eqref{eq:cov_deriv_gamma} reads after evaluating
the Dirac matrix commutators,
\begin{align}
 \nabla_{\mu} \gamma^{\nu} = \left[ \Delta \Gamma_{\mu} , \gamma^{\nu} \right] + K^{\nu}_{\point \mu \kappa} \gamma^{\kappa}.
\label{eq:gammadertorsion1}
\end{align}
Using the constraint \eqref{eq:def_Delta_Gamma}, this implies
\begin{align}
 \nabla_{\mu} \gamma^{\mu} = K^{\mu}_{\point \mu \kappa} \gamma^{\kappa}.
\label{eq:gammadertorsion2}
\end{align}
In the presence of torsion with $K^{\kappa}_{\point \kappa \mu} \neq
0$, this result is incompatible with the vierbein postulate
\eqref{eq:vierbein_postulate}. In the present notation, the latter is
equivalent to the vanishing covariant derivative of the Dirac matrices
\begin{align}
 {\nabla_\ttm{(\vierbein)}}_{\mu} {\gamma_\ttm{(\vierbein)}}^{\nu} = 0 \text{,}
\label{eq:vierbein_postulate2}
\end{align}
where the ${\gamma_\ttm{(\vierbein)}}^{\nu}$ follow the vierbein
construction \Eqref{eq:gammae}. This discrepancy between our more
general formalism and the conventional vierbein formalism is in line
with the fact that the inclusion of torsion requires to go beyond the
conventional vierbein formalism such as, e.g., Einstein-Cartan
theory. From the viewpoint of our spin-base invariant formalism,
torsion can be accommodated in a straightforward manner on the basis
of our requirements of Sect.~\ref{sec:Axioms}.\footnote{It is
  worthwhile to note that Eqs. \eqref{eq:gammadertorsion1} and
  \eqref{eq:gammadertorsion2} actually do not intertwine torsion and
  the Dirac matrices. Since torsion is naturally contained in the full
  covariant derivative on the left-hand side,
  cf. Eqs. \eqref{def:gencov} and \eqref{def:spacetimeconnection}, the
  torsion terms naturally drop out of
  Eqs. \eqref{eq:gammadertorsion1} and \eqref{eq:gammadertorsion2}. By
  contrast, torsion is trivially constrained to vanish in
  \Eqref{eq:vierbein_postulate2}, if the covariant derivative on the
  left-hand side is assumed to also contain the antisymmetric part of
  the spacetime affine connection. Hence, it seems that the vierbein
  formalism could also accommodate torsion, if the vierbein postulate
  is generalized analogously to \Eqref{eq:gammadertorsion1}.}

In the remainder of this section, we concentrate on the analysis of
the properties of the $\Delta \Gamma_\mu$ part of the spin
connection. For this, it is useful to explicitly construct the spin
metric $h$, as is done in App. \ref{App:spin_metric}. For a given set
of Dirac matrices, the spin metric turns out to be uniquely fixed up
to a sign and can be parametrized by
\begin{align}
 h ={}& \pm \cplx \euler^{\cplx \frac{\varphi}{2}} \euler^{\hat{M}}, \label{eq:spin_metric}
\end{align}
where $\varphi$ and $\hat{M}$ are (up to a sign) implicitely defined
by
\begin{align}
 \gamma_{\mu}^{\dagger} = - \euler^{\hat{M}} \gamma_{\mu} \euler^{-\hat{M}} \text{,} \quad \tr \hat{M} = 0 \text{,} \quad \euler^{\hat{M}^{\dagger}} = \euler^{\cplx \varphi} \euler^{\hat{M}}.
\end{align}
The angle $\varphi$ can only take discrete constant values,
\begin{align}
 \varphi \in \left\{ n \frac{2 \pi}{d_{\gamma}} : n \in \{ 0, \ldots, d_{\gamma} - 1 \} \right\} , \quad \partial_{\mu} \varphi = 0 \text{.} \label{eq:spin_metric2}
\end{align}
These properties together with \Eqref{eq:cov_deriv_spin_metric} imply
another constraint for the spin connection (see
App. \ref{App:spin_metric} for details):
\begin{align}
 \Gamma_{\mu} + \bar{\Gamma}_{\mu} = \hat{\Gamma}_{\mu} + \bar{\hat{\Gamma}}_{\mu} = h^{-1} \partial_{\mu} h \text{.} \label{eq:Gamma_real}
\end{align}
Even if we admit for a nonzero trace of the spin connection
(cf. Sect. \ref{sec:gauge_field}), this constraint together with the
properties of the spin metric imply that the trace part has to be
purely imaginary
\begin{align}
 \Re \tr \Gamma_{\mu} = 0.
\end{align}
If we span $\Delta \Gamma_{\mu}$ also by the standard Clifford basis
\begin{align}\label{eq:decomp_Gamma}
 \Delta \Gamma_{\mu} = \Delta p_{\mu} \gamma_{\ast} + \Delta v_{\mu}^{\point \alpha} \gamma_{\alpha} + \Delta a_{\mu}^{\point \alpha} \gamma_{\ast} \gamma_{\alpha} + \Delta t_{\mu}^{\point \alpha \beta} [ \gamma_{\alpha} , \gamma_{\beta} ]
\end{align}
and use the constraints \eqref{eq:def_Delta_Gamma} and (\ref{eq:Gamma_real}), we conclude that
\begin{align}
\begin{aligned}
 \rmi \quad{}& \Delta p_{\mu} = 0 \text{,}\\
 \rmii \quad{}& \Delta v_{[\alpha \beta]} = 0 \text{,} \quad \Delta v_{\mu}^{\point \alpha} \in \R \text{,}\\
 \rmiii \quad{}& \Delta a_{\alpha}^{\point \alpha} = 0 \text{,} \quad \Delta a_{\mu}^{\point \alpha} \in \R \text{,}\\
 \rmiv \quad{}& \Delta t_{\mu}^{\point (\alpha \beta)} = 0 \text{,} \quad \Delta t_{\beta}^{\point \beta \alpha} = 0 \text{,} \quad \Delta t_{\mu}^{\point \alpha \beta} \in \R \text{,}
\end{aligned}
\label{eq:DeltaCoeffs}
\end{align}
where we use $(...)$ to denote complete symmetrization of indices. In
summary, this leaves us with 45 real parameters for
$\Delta\Gamma_\mu$. It is important to note that the coefficient
tensors in \Eqref{eq:DeltaCoeffs} do not change under spin base
transformations, since
\begin{align}
 \Delta \Gamma_{\mu} \rightarrow \mcS \Delta \Gamma_{\mu} \mcS^{-1} \text{,}
\end{align}
transforms homogeneously in contrast to $\Gamma_\mu$ and
$\hat{\Gamma}_\mu$, cf. \Eqref{eq:gauge_trafo_Gamma}. Hence, spin-base
transformations cannot be employed to transform any of these
parameters to zero.

We interpret $\Delta \Gamma_{\mu}$ as a \textit{spin torsion}. Similarly to
general relativity where the torsion becomes visible in the
antisymmetric part of $\Gamma^{\lambda}_{\mu \nu}$, also $\Delta
\Gamma_{\mu}$ is contained in the \textit{antisymmetric} part of
the affine connection $\Gamma_{\mu}$, cf. \Eqref{eq:Gamma_real}
\begin{align}
 \frac{1}{2} (\Gamma_{\mu} - \bar{\Gamma}_{\mu}) = \frac{1}{2} (\hat{\Gamma}_{\mu} - \bar{\hat{\Gamma}}_{\mu}) + \Delta \Gamma_{\mu}\text{,}
\end{align}
where anti-symmetrization is defined in terms of Dirac conjugation.
Also the transformation behavior is reminiscent to that of torsion, since it
transforms homogeneously under spin-base transformations and coordinate
transformations.

In order to illustrate the physical meaning of $\Delta \Gamma_\mu$,
let us consider the contribution of this spin torsion to the Dirac
operator. Using the identities (valid for $d = d_{\gamma} = 4$),
\begin{align}
 \gamma_{\ast} [\gamma^{\alpha} , \gamma^{\beta} ] ={}& \frac{\cplx}{2} \tilde{\varepsilon}^{\alpha \beta \mu \nu} [\gamma_{\mu} , \gamma_{\nu}]\\
 \{ \gamma^{\mu} , [\gamma^{\alpha} , \gamma^{\beta}] \} ={}& - 4 \cplx \tilde{\varepsilon}^{\mu \alpha \beta \nu} \gamma_{\ast} \gamma_{\nu} \text{,}
\end{align}
and taking the constraints \eqref{eq:DeltaCoeffs} into account,
we find
\begin{equation}
 \bar{\psi} \gamma^{\mu} \Delta \Gamma_{\mu} \psi = \mathscr{M} \bar{\psi} \psi - \mathscr{A}_{\mu} \bar{\psi} \cplx \gamma_{\ast} \gamma^{\mu} \psi - \mathscr{F}_{\mu \nu} \bar{\psi} \frac{\cplx}{4} [\gamma^{\mu} , \gamma^{\nu} ] \psi \text{,}
\label{eq:DiracOpSpinT}
\end{equation}
Here we have introduced the intuitive abbreviations
\begin{align}
 \mathscr{M} ={}& \Delta v_{\alpha}^{\point \alpha} \\
 \mathscr{A}^{\nu} ={}& 2 \Delta t_{\mu \alpha \beta} \tilde{\varepsilon}^{\mu \alpha \beta \nu} \\
 \mathscr{F}_{\mu \nu} ={}& \Delta a^{[\alpha \beta]} \tilde{\varepsilon}_{\alpha \beta \mu \nu} \text{,}
\end{align}
for a scalar field (spacetime dependent mass) $\mathscr{M}$, an axial
vector field $\mathscr{A}^{\nu}$, and an antisymmetric tensor field
$\mathscr{F}_{\mu \nu}$ all of which have mass dimension one. We
conclude that such fields can be accommodated in the spin
torsion. They can obviously remain nonzero even in the limit of flat
Minkowski space. It is interesting to observe that no pseudo-scalar
and no vector field, which would complete the possible bilinear
fermion structures, occur in \Eqref{eq:DiracOpSpinT}. As discussed in
Sect.~\ref{sec:gauge_field}, a vector field can straightforwardly be
accommodated in the trace part of the spin connection.

Out of the 45 parameters of the spin torsion, the fields
$\mathscr{M}$, $\mathscr{A}^{\nu}$, and $\mathscr{F}_{\mu \nu}$
summarize 11 parameters. The remaining 34 can contribute to higher
order operators, e.g., involving more covariant derivatives.

This comparatively large number of parameters of the spin torsion can,
of course, be further constrained by additional symmetry
requirements. For instance, in order to construct a chiral symmetry,
we demand for a covariantly constant $\gamma_\ast$ which facilitates
the construction of covariantly constant chiral projectors,
\begin{align}
 0 = \slashed{\nabla} \gamma_{\ast} = \gamma^{\mu} [ \Delta \Gamma_{\mu} , \gamma_{\ast} ] = 2 \Delta v_{\mu}^{\point \mu} \gamma_{\ast} - \Delta a_{\mu \nu} [ \gamma^{\mu} , \gamma^{\nu} ]\text{,}
\end{align}
which implies additional constraints for the spin torsion
\begin{align}\label{eq:add_constraints}
 \Delta v_{\mu}^{\point \mu} = 0 \text{,} \quad \Delta a_{[\mu \nu]} = 0 \text{,}
\end{align}
leaving 38 free real parameters. In fact, this chiral-symmetry
constraint requires only the scalar field $\mathscr{M}$ and the
antisymmetric tensor field $\mathscr{F}_{\mu \nu}$ to vanish. The
axial vector field $\mathscr{A}^{\nu}$ (4 parameters) as well as the
remaining 34 parameters possibly contributing to higher order
operators are left untouched.

To summarize the present section, we now have a covariant derivative of
Dirac fermions at our disposal which encodes a parallel transport of a
Dirac spinor in curved spacetimes that respects general coordinate
invariance as well as local spin-base invariance. More explicitly, given a spinor
$\psi$ which transforms as a scalar under coordinate transformations
and a vector under spin-base transformations, its covariant derivative can be written as
\begin{align}
 \nabla_{\mu} \psi = \partial_{\mu} \psi + \hat{\Gamma}_{\mu} \psi + \Delta \Gamma_{\mu} \psi,
\label{eq:spinCD}
\end{align}
where $\hat{\Gamma}_\mu$ is fully determined in terms of spacetime
dependent Dirac matrices also carrying metric information and
$\Delta{\Gamma}_\mu$ denotes the spin torsion. This is in complete
analogy to the covariant derivative of a spacetime vector which can be
written in terms of the Levi-Civita connection (determined in terms of
the metric) and the spacetime torsion. We would like to emphasize that
the spin torsion and the spacetime torsion are mutually
independent. They have to be fixed by corresponding external
conditions or a corresponding dynamical theory.

\section{Lie derivative}
\label{sec:lie}

The standard Lie derivative $\mcL_v$ with respect to a vector
field $v^\mu$ (considered as infinitesimal in the following) is defined by
\begin{align}
 \mcL_{v_{1}} {v_{2}}^{\mu} = {v_{1}}^{\nu} \partial_{\nu} {v_{2}}^{\mu} - {v_{2}}^{\nu} \partial_{\nu} {v_{1}}^{\mu} \text{,}
\end{align}
where ${v_2}^\mu$ is also some vector field. This geometrical
structure can be used to implement the statement that Einstein's
theory of general relativity is torsion-free. Demanding that
the Lie derivative also equals the covariantized right-hand side,
\begin{align}
\mcL_{v_{1}} {v_{2}}^{\mu}\, \stackrel{!}{=}\, {v_{1}}^{\nu}
  D_{\nu} {v_{2}}^{\mu} - {v_{2}}^{\nu} D_{\nu} {v_{1}}^{\mu},
\label{eq:elimtorsion}
\end{align}
the torsion has to vanish.

This relation implies that $\Gamma^{\lambda}_{\mu \nu}$ has to be
symmetric in $\mu \leftrightarrow \nu$ and therefore is equal to the
Levi-Civita connection. If we wish to apply the same concept to the
spin-base covariant derivative in order to exclude spin torsion, we
first need a Lie derivative for spinors. In fact this has been a
challenge of its own which has been extensively discussed in the
literature \cite{Kosmann:1966,Fatibene:1996tf,Sharipov:2008xv}.

In the following, we present an independent definition of a
generalized Lie derivative for spinors $\tilde{\mcL}$ which is
motivated by the Weldon theorem \Eqref{eq:variation_gamma}.  Since the
metric is encoded in the Dirac matrices in the present spin-base
invariant formalism, it is natural to define the generalized Lie
derivative in terms of its action on the Dirac matrices. From the
Weldon theorem \Eqref{eq:variation_gamma}, we know that general
Clifford-algebra compatible variations of the Dirac matrices can be
decomposed into a metric variation $\delta \metric^{\mu \nu}$ and an
infinitesimal spin-base transformation $\delta \mcS_{\gamma}$. As the
Lie derivative can be related to infinitesimal diffeomorphisms, the
variation of the metric occuring in the Weldon theorem
\Eqref{eq:variation_gamma} is given by the ordinary Lie derivative
\begin{align}
 \delta \metric^{\mu \nu} = \mcL_{v} \metric^{\mu \nu} = - g^{\mu \rho} \partial_{\rho} v^{\nu} - g^{\nu \rho} \partial_{\rho} v^{\mu} + v^{\rho} \partial_{\rho} g^{\mu \nu} \text{.}
\end{align}
However, in order to compare spinors under a variation
of the metric without contributions from local spin-base variations,
we keep the spin bases fixed. Hence, we define the generalized Lie derivative in terms of a variation of the Dirac matrices with  $\delta \mcS_{\gamma}=0$
\begin{align}\label{eq:def_Lie_derivative}
 \tilde{\mcL}_{v} \gamma^{\mu} = \frac{1}{2} (\mcL_{v} g^{\mu \nu}) \gamma_{\nu} \text{.}
\end{align}
This way the Lie derivative gives us the variation of the spinors
under diffeomorphisms with fixed spin bases, corresponding to a
\textit{comparability} of the in general different spin bases under
\textit{different} metrics. Of course we also demand the generalized Lie
derivative $\tilde{\mcL}$ to fulfill a product rule and to coincide
with the ordinary Lie derivative $\mcL$ if the considered object is a
scalar under spin-base transformations,
\begin{align}
 \tilde{\mcL}_{v} \bar{\psi} \psi = \mcL_{v} \bar{\psi} \psi \text{.}
\end{align}
That justifies the general form of $\tilde{\mcL}$:
\begin{align}
\begin{aligned}
 \tilde{\mcL}_{v} \psi ={}& \mcL_{v} \psi + \mcZ_{v} \psi \text{,}\\
 \tilde{\mcL}_{v} \bar{\psi} ={}& \mcL_{v} \bar{\psi} - \bar{\psi} \mcZ_{v} \text{,} \\
 \tilde{\mcL}_{v} \gamma^{\mu} ={}& \mcL_{v} \gamma^{\mu} + [ \mcZ_{v} , \gamma^{\mu} ]
\end{aligned}
\end{align}
for some Clifford-algebra valued matrix $\mcZ_{v}$. We demand
additionally for $\mcZ_{v}$ to be traceless,
\begin{align}\label{eq:vanishing_tr_mcZ}
 \tr \mcZ_{v} = 0\text{.}
\end{align}
This condition is natural, as any nonzero trace part $\sim \mrI$ would
not modify the Dirac matrices and hence leave also the geometry
unaffected. Even if $\mcZ_{v}$ was not traceless, the trace part would
act similar to that of the covariant derivative discussed in the next
section and hence carry no independent information.  The traceless
part of $\mcZ_{v}$ can be calculated from
\Eqref{eq:def_Lie_derivative}, by a comparison with the ordinary Lie
derivative of the Dirac matrices which can be derived
straightforwardly,
\begin{align}
 \mcL_{v} \gamma^{\mu} = \frac{1}{2} ( \mcL_{v} g^{\mu \nu}) \gamma_{\nu} - \! \left[ v^{\rho} \hat{\Gamma}_{\rho} + \frac{1}{8} (\partial_{[\rho} v_{\lambda]}) [ \gamma^{\rho} , \gamma^{\lambda}] , \gamma^{\mu} \right].
\end{align}
Hence, we can read off
\begin{align}
 \mcZ_{v} = v^{\rho} \hat{\Gamma}_{\rho} + \frac{1}{16} (\partial_{\rho} v_{\lambda} - \partial_{\lambda} v_{\rho}) [ \gamma^{\rho} , \gamma^{\lambda}] \text{.}
\end{align}
This line of argument leads us to a generalized Lie derivative for
Dirac spinors
\begin{align}
 \tilde{\mcL}_{v} \psi = v^{\rho} \partial_{\rho} \psi + v^{\rho} \hat{\Gamma}_{\rho} \psi + \frac{1}{8} (\partial_{[\rho} v_{\lambda]}) [\gamma^{\rho} , \gamma^{\lambda}] \psi \text{.}
\end{align}
Now, the geometric argument for eliminating the spin torsion analogous
to that of general relativity formulated by \Eqref{eq:elimtorsion} can
be completed: the analog requirement in spinor space is to demand our
generalized Lie derivative to agree with a spinor-covariantized form:
\begin{align}\label{eq:cov_lie_derivative}
 \tilde{\mcL}_{v} \psi = v^{\rho} \nabla_{\rho} \psi + \frac{1}{8} (\partial_{[\rho} v_{\lambda]} ) [ \gamma^{\rho} , \gamma^{\lambda} ] \psi \text{.}
\end{align}
Then we can immediately conclude that
\begin{align}
 v^{\rho} \Delta \Gamma_{\rho} = 0 \text{,}
\end{align}
for all (infinitesimal) vectors $v^{\rho}$. Therefore, relating the
geometrical construction represented by the Lie derivative to the
covariant derivative in spinor space implies that the spin torsion has
to vanish.

In \Eqref{eq:cov_lie_derivative}, we have only covariantized the
spinorial part.  Alternatively, we could also require the generalized
Lie derivative to agree with its fully covariantized form leading to
\begin{align}
 \tilde{\mcL}_{v} \psi = v^{\rho} \nabla_{\rho} \psi + \frac{1}{8}
 (D_{[\rho} v_{\lambda]} ) [\gamma^{\rho}, \gamma^{\lambda}] \psi.
\label{eq:covLie}
\end{align}
This requirement relates torsion and spin torsion,
\begin{align}
 \Delta \Gamma_{\mu} = \frac{1}{8} C_{\mu \rho \lambda} [\gamma^{\rho} , \gamma^{\lambda}] \text{,} \quad C^{\sigma}_{\point \sigma \mu} = 0 \text{.}
\end{align}
Read together with Eqs.~$\rmv$ of \eqref{eq:spintorsionconstraints}
and \eqref{eq:decomp_Gamma}, this resembles the form of the spin
connection $\Gamma_{\mu} = \hat{\Gamma}_{\mu} + \Delta \Gamma_{\mu}$,
known from the vierbein formalism (with torsion replaced by contorsion), but with the additional constraint
that the space time torsion needs to be traceless.

However, it is important to emphasize that this relation between
  spin torsion and torsion is only nontrivial, as long as we do not
  impose the condition \eqref{eq:elimtorsion} for Lie derivatives of
  vectors. In fact, treating spinors and vectors differently appears
  unnatural. Hence, imposing the covariantized form also for space
time vectors, we have
\begin{align}
 \tilde{\mcL}_{v} (\bar{\psi} \gamma^{\nu} \psi) = \mcL_{v} (\bar{\psi} \gamma^{\nu} \psi) = v^{\rho} D_{\rho} (\bar{\psi} \gamma^{\mu} \psi) - (\bar{\psi} \gamma^{\rho} \psi) D_{\rho} v^{\mu}
\end{align}
which in combination with \Eqref{eq:covLie} immediately
  implies that both kinds of torsion have to vanish
\begin{align}
\begin{aligned}
 \rmi {}& \quad C^{\rho}_{\point \mu \nu} = 0, \\
 \rmii {}& \quad \Delta \Gamma_{\mu} = 0 \text{.}
\end{aligned}
\end{align}
The fully covariantized form for our generalized
Lie derivative in \Eqref{eq:covLie} is identical (up to torsion) to the Kosmann-Lie
derivative discussed in the literature
\cite{Kosmann:1966,Fatibene:1996tf,Sharipov:2008xv}.

\section{Gauge Fields}
\label{sec:gauge_field}

In the preceding sections, we have set a possible trace part of the
spin connection $\Gamma_{\mu}$ to zero, as such a trace part
proportional to the identity in Dirac space $\sim \mrI$ does not
transform the Dirac matrices nontrivially,
cf. \Eqref{eq:cov_deriv_gamma}. If we allow for this generalization,
the symmetry group can be extended to $\mcG \otimes \text{SL}(d_{\gamma},
\C)$, where $\mcG$ denotes the symmetry group of the trace part.  The
Clifford algebra is, of course, also invariant under this larger
group, since the Dirac matrices and thus the geometry do not transform
under $\mfrg\in\mcG$.

To construct a connection ${\Gamma_\ttm{(\mcG \! \otimes \!
    \text{SL})}}_{\mu}$ for this larger group, we consider symmetry
transformations $\mfrg \! \otimes \! \mcS \in \mcG \otimes
\text{SL}(d_{\gamma} , \C)$ and find analogously to
\Eqref{eq:gauge_trafo_Gamma}
\begin{align}
 {\Gamma_\ttm{(\mcG \! \otimes \! \text{SL})}}_{\mu} \! 
\rightarrow \! \mfrg \! \otimes \! \mcS \, {\Gamma_\ttm{(\mcG \! \otimes \! \text{SL})}}_{\mu} (\mfrg \! \otimes \! \mcS)^{-1} \! 
- \! \big( \partial_{\mu} (\mfrg \! \otimes \! \mcS) \big) (\mfrg \! \otimes \! \mcS)^{-1}
\end{align}
as the transformation property of the spin connection. Here we can use
the product rule for the derivative and expand the inhomogenous part,
\begin{align}
 \big( \partial_{\mu} (\mfrg \! \otimes \! \mcS) \big) (\mfrg \! \otimes \! \mcS)^{-1} \! = \! \big( (\partial_{\mu} \mfrg) \mfrg^{-1} \big) \! \otimes \! \mrI \! + \! \mrI_\ttm{(\mcG)} \! \otimes \! \big( (\partial_{\mu} \mcS) \mcS^{-1} \big) \text{,}
\end{align}
where $\mrI_\ttm{(\mcG)}$ is the unit element of $\mcG$. Because of
this behavior, it is sufficient to consider connections with the property
\begin{align}
 {\Gamma_\ttm{(\mcG \! \otimes \! \text{SL})}}_{\mu} = {\Gamma_\ttm{(\mcG)}}_{\mu} \! \otimes \! \mrI + \mrI_\ttm{(\mcG)} \! \otimes \! \Gamma_{\mu} \text{,}
\label{eq:gaugespinconnection}
\end{align}
where ${\Gamma_\ttm{(\mcG)}}_{\mu}$ is the connection for the group
$\mcG$ and $\Gamma_{\mu}$ is the traceless connection for the
$\text{SL}(d_{\gamma},\C)$ part, i.e. the $\Gamma_{\mu}$ from the previous
sections. Obviously, the Dirac trace part of ${\Gamma_\ttm{(\mcG \!
    \otimes \! \text{SL})}}_{\mu}$ accommodates the connection for the group
$\mcG$.

Similarly, a straightforward generalization of the spin metric
suggests the form
\begin{align}
 h_\ttm{(\mcG \! \otimes \! \text{SL})} = \mrI_\ttm{(\mcG)} \! \otimes \! h \text{,}
\label{eq:gaugespinmetric}
\end{align}
with the corresponding transformation law
\begin{align}
 h_\ttm{(\mcG \! \otimes \! \text{SL})} \rightarrow (\mfrg^{\dagger} \! \otimes \! \mcS^{\dagger})^{-1} h_\ttm{(\mcG \! \otimes \! \text{SL})} (\mfrg \! \otimes \! \mcS)^{-1}
\label{eq:blabla1}
\end{align}
under a $\mfrg \otimes \mcS$ transformation. Requiring the
transformation \eqref{eq:blabla1} to preserve
\Eqref{eq:gaugespinmetric}, the elements of $\mcG$ need to be unitary
\begin{align}
 \mfrg^{-1} = \mfrg^{\dagger} \text{.}
\end{align}
If we now additionally demand for metric compatibility,
\Eqref{eq:cov_deriv_spin_metric}, we get
\begin{align}
 \mrI_{(\mcG)} \! \otimes \! \big( h^{-1} ( \partial_{\mu} h) \big) = {\Gamma_\ttm{(\mcG \! \otimes \! \text{SL})}}_{\mu} + \mrI_\ttm{(\mcG)} \! \otimes \! h^{-1} \, {\Gamma_\ttm{(\mcG \! \otimes \! \text{SL})}^{\dagger}}_{\mu} \, \mrI_\ttm{(\mcG)} \! \otimes \! h \text{,}
\end{align}
from which we deduce with regard to \Eqref{eq:gaugespinconnection}
that the connection of $\mcG$ needs to be antihermitean
\begin{align}
 {\Gamma_\ttm{(\mcG)}^{\dagger}}_{\mu} = - {\Gamma_\ttm{(\mcG)}}_{\mu} \text{.}
\end{align}
Here we also used \Eqref{eq:Gamma_real}. This justifies to introduce
the gauge field $\mcA_{\mu}$
\begin{align}
 {\Gamma_\ttm{(\mcG)}}_{\mu} = \cplx \mcA_{\mu}
\end{align}
which is associated with the $\mcG$ symmetry. This field can in
general be non-abelian but is always hermitean as is conventional in
ordinary gauge field theory.

To summarize, the inclusion of a trace part in the spin connection
$\Gamma_\mu$ can be viewed as an extension of the symmetry group from
$\text{SL}(d_{\gamma} , \C)$ to $\mcG \otimes \text{SL}(d_{\gamma}, \C)$, with
$\mcG$ being a unitary group. The spin connection can then be decomposed as
\begin{align}
 {\Gamma_\ttm{(\mcG \! \otimes \! \text{SL})}}_{\mu} = \cplx \mcA_{\mu} \! \otimes \! \mrI + \mrI_\ttm{(\mcG)} \! \otimes \! (\hat{\Gamma}_{\mu} + \Delta \Gamma_{\mu})\text{,}
\end{align}
or in short
\begin{align}
 \Gamma_{\mu} = \cplx \mcA_{\mu} + \hat{\Gamma}_{\mu} + \Delta \Gamma_{\mu} \text{,}
\end{align}
as it is understood and used in the following. Within the physical
context of fermions in curved space, the $\text{SL}(d_{\gamma} , \C)$ part of
the connection is always present in covariant derivatives of spinor
fields, since it carries the information about how fermions evolve
dynamically in a given curved space. By contrast, the gauge part of
the connection may or may not be present depending on whether a
fermion is charged under the group $\mcG$. Technically, the
distinction among differently charged fermions may be parametrized by
a charge matrix as a factor inside $\mcA_{\mu}$.

\section{Spin curvature}
\label{sec:spin_curv}

From our knowledge about the spinor convariant derivative and the
associated spin connection, it is immediate to construct a curvature
or field strength which we denote by spin curvature for short. Again,
we motivate the definition for this spin curvature by analogy to the
standard definition of the curvature tensor in general relativity
(including torsion) \cite{Watanabe:2004nt},
\begin{align}
 R_{\mu \nu \point \rho}^{\point[2] \lambda} T^{\rho} = [ D_{\mu} , D_{\nu} ] T^{\lambda} + C^{\sigma}_{\point \mu \nu} D_{\sigma} T^{\lambda} \text{,} \quad \forall T^{\rho} \text{ tensor.}
\label{eq:Riemann}
\end{align}
This suggests the definition of the spin curvature $\Phi_{\mu \nu}$,
\begin{align}
 \Phi_{\mu \nu} \psi = [ \nabla_{\mu} , \nabla_{\nu} ] \psi + C^{\sigma}_{\point \mu \nu} \nabla_{\sigma} \psi \text{.}
\end{align}
More explicitly, it is given by
\begin{align}
 \Phi_{\mu \nu} \! ={}& \partial_{\mu} \Gamma_{\nu} - \partial_{\nu} \Gamma_{\mu} + [ \Gamma_{\mu} , \Gamma_{\nu}] \\
 ={}& \! \cplx \mcF_{\mu \nu} \! + \! \hat{\Phi}_{\mu \nu} \! + \! 2 \partial_{[\mu} \Delta \Gamma_{\nu]} \! + \! 2 [\hat{\Gamma}_{[\mu} , \Delta \Gamma_{\nu]}] \! + \! [ \Delta \Gamma_{\mu} , \Delta \Gamma_{\nu} ] \text{,}
\label{eq:spincurvature}
\end{align}
where $\mcF_{\mu \nu}$ is the field strength tensor of the gauge field
\begin{align}
 \mcF_{\mu \nu} = \partial_{\mu} \mcA_{\nu} - \partial_{\nu} \mcA_{\mu} + [\mcA_{\mu} , \mcA_{\nu}].
\end{align}
The quantity $\hat{\Phi}_{\mu \nu}$ is the spin curvature induced by $\hat{\Gamma}_{\mu}$
\begin{align}
 \hat{\Phi}_{\mu \nu} = \partial_{\mu} \hat{\Gamma}_{\nu} - \partial_{\nu} \hat{\Gamma}_{\mu} + [ \hat{\Gamma}_{\mu} , \hat{\Gamma}_{\nu} ] \text{,}
\end{align}
which can be related to the curvature tensor ${R_\ttm{(\mathrm{LC})}}_{\mu
  \nu \point \rho}^{\point[2] \lambda}$ defined in terms of the
Levi-Civita connection ${D_\ttm{(\mathrm{LC})}}_{\mu}$ by the following observation:
\begin{align}
 {R_\ttm{(\mathrm{LC})}}_{\mu \nu \point \rho}^{\point[2] \lambda}
 \gamma^{\rho} =
       [{D_\ttm{(\mathrm{LC})}}_{\mu},{D_\ttm{(\mathrm{LC})}}_{\nu}]
       \gamma^{\lambda} = - [ \hat{\Phi}_{\mu \nu} , \gamma^{\lambda}
       ] \text{.}
\end{align}
This demonstrates the correspondence between the spin-torsion/torsion
free curvature expressions:
\begin{align}
 \hat{\Phi}_{\mu \nu} = \frac{1}{8} {R_\ttm{(\mathrm{LC})}}_{\mu \nu \lambda \rho} [ \gamma^{\lambda} , \gamma^{\rho}] \text{,}
\end{align}
where 
\begin{align}
 {R_\ttm{(\mathrm{LC})}}_{\mu \nu \point \rho}^{\point[2] \lambda} = \partial_{\mu} \christoffel{\lambda \\ \nu \rho} \! - \! \partial_{\nu} \christoffel{\lambda \\ \mu \rho} \! + \! \christoffel{\lambda \\ \mu \sigma} \christoffel{\sigma \\ \nu \rho} \! - \! \christoffel{\lambda \\ \nu \sigma} \christoffel{\sigma \\ \mu \rho} \text{.}
\end{align}
These results together with the explicit representation
\eqref{eq:spincurvature} make it clear that the spin curvature does
not carry information about the whole spacetime structure, since no
spacetime-torsion dependent terms occur. Only spin torsion appears in
\Eqref{eq:spincurvature} which is in line with the fact that also a
covariant derivative acting on a Dirac spinor does not depend on
spacetime torsion but only on spin torsion, cf. \Eqref{eq:spinCD}. A
direct coupling between spinor degrees of freedom and spacetime
torsion therefore requires ad-hoc higher-order coupling terms or
higher-spin fields such as Rarita-Schwinger spinors, see below.

As a simple application of the spin curvature, let us construct the
simplest classical field theory that can be formed out of the spin
curvature. Since, $\Phi_{\mu\nu}$ is Clifford-algebra valued, there
exists already a spin-base and diffeomorphism invariant quantity to
linear order in the spin curvature. The simplest classical action thus
is
\begin{align}
 S_{\Phi} = \frac{1}{16\pi G} \regint{x} \mfrL_{\Phi} \text{,} \quad \mfrL_{\Phi} = \frac{1}{{d_{\mcA}} d_{\gamma}} \tr (\gamma^{\mu} \Phi_{\mu \nu} \gamma^{\nu}) \text{,} \label{eq:SPhi}
\end{align}
where $G$ is a coupling constant, and $d_{\mcA}$ is the
  dimension of the representation of the gauge group $d_{\mcA}=\tr
  \mrI_\ttm{(\mcG)}$. We set $d_{\mcA}=1$ in the absence of any gauge
  group. The content of this field theory can be worked out more
explicitly, using the identities
\begin{align}
 {D_\ttm{(\mathrm{LC})}}_{\mu} \Delta \Gamma_{\nu} ={}& ({D_\ttm{(\mathrm{LC})}}_{\mu} \Delta v_{\nu}^{\point \alpha}) \gamma_{\alpha} + ({D_\ttm{(\mathrm{LC})}}_{\mu} \Delta a_{\nu}^{\point \alpha}) \gamma_{\ast} \gamma_{\alpha} \notag\\
 {}& + ({D_\ttm{(\mathrm{LC})}}_{\mu} \Delta t_{\nu}^{\point \alpha \beta}) [\gamma_{\alpha} , \gamma_{\beta} ] - [ \hat{\Gamma}_{\mu} , \Delta \Gamma_{\mu}] \text{,}\\
 0 = \tr([\gamma^{\alpha},\gamma^{\beta}{}&]) = \tr(\gamma^{\mu} [\gamma^{\alpha},\gamma^{\beta}]) = \tr(\gamma_{\ast} \gamma^{\mu} [\gamma^{\alpha}, \gamma^{\beta}]) \text{,}\\
 \tr([\gamma^{\mu},\gamma^{\nu}][\gamma^{\alpha},{}&\gamma^{\beta}]) = 4 d_{\gamma} (\metric^{\mu \beta} \metric^{\nu \alpha} - \metric^{\mu \alpha} \metric^{\nu \beta}).
\end{align}
The Lagrangian reads in terms of the Levi-Civita curvature and the
spin torsion coefficients
\begin{align}
 \mfrL_{\Phi} ={}& \frac{1}{2} R_\ttm{(\mathrm{LC})} + 2 (\Delta v_{\mu}^{\point \mu})^2 - 2 \Delta v_{\mu}^{\point \nu} \Delta v_{\nu}^{\point \mu} \notag \\
 {}& + 2 \Delta a_{\mu}^{\point \nu} \Delta a_{\nu}^{\point \mu} + 32 \Delta t_{\mu \nu \kappa} \Delta t^{\kappa \nu \mu} \text{,}
\label{eq:spinlagrangian}
\end{align}
where $R_\ttm{(\mathrm{LC})} = {R_\ttm{(\mathrm{LC})}}_{\mu
  \nu}^{\point[2] \mu \nu}$.  This action is rather similar to the
(torsion-amended) Einstein-Hilbert action
\begin{align}
 S_\ttm{R} ={}& \frac{1}{32 \pi G} \regint{x} R \text{,} \quad R = R_{\mu \nu}^{\point[2] \mu \nu} \text{,} \label{eq:SR}\\
 R ={}& R_\ttm{(\mathrm{LC})} + 2 {D_\ttm{(\mathrm{LC})}}_{\mu} K^{\mu \point \nu}_{\point \nu} - K^{\rho \point \nu}_{\point \nu} K^{\point[2] \mu}_{\rho \mu} + K^{\rho \mu \nu} K_{\rho \nu \mu}, \notag
\end{align}
which differs from \Eqref{eq:spinlagrangian} only in the torsion
terms. Obviously, $ \mfrL_{\Phi}$ cannot depend on spacetime torsion
as $\Phi$ is blind to spacetime torsion as well. 

This simple observation offers a speculative though interesting
perspective: if classical GR was based on \Eqref{eq:SPhi} (and possibly
supplemented by higher order monomials of $\Phi_{\mu\nu}$) instead of
\Eqref{eq:SR}, the absence of spacetime torsion in classical GR would
be a natural self-evident consequence.

Instead, $S_\Phi$ confronts us with the presence of spin torsion terms
in \Eqref{eq:spinlagrangian}. In this simplest field theory, however,
the spin torsion terms occur only algebraically, implying that the
torsion fields remain non-dynamical and satisfy particularly simple
equations of motion.

Varying the action with respect to the fields $\Delta v_{\mu}^{\point
  \nu}$, $\Delta a_{\mu}^{\point \nu}$ and $\Delta t_{\mu}^{\point
  \alpha \beta}$, taking into account the constraints
\eqref{eq:spintorsionconstraints}, we find
\begin{align}
 \delta \Delta v^{\mu \nu} \! ={}& \!\! \frac{1}{2} (\delta^{\mu}_{\rho} \delta^{\nu}_{\lambda} + \delta^{\nu}_{\rho} \delta^{\mu}_{\lambda}) \, \delta \Delta v^{\rho \lambda} \text{,}\\
 \delta \Delta a^{\mu \nu} \! ={}& \!\! \left[ \delta^{\mu}_{\rho} \delta^{\nu}_{\lambda} - \frac{1}{4} \metric^{\mu \nu} \metric_{\rho \lambda} \right] \! \delta \Delta a^{\rho \lambda} \text{,}\\
 \delta \Delta t^{\mu \alpha \beta} \! ={}& \!\! \left[ \! \delta_{\rho}^{\mu} \delta_{\lambda}^{[\alpha} \delta_{\sigma}^{\beta]} \! - \! \frac{1}{3} \metric_{\rho \lambda} \metric^{\mu [\alpha} \delta_{\sigma}^{\beta]} \! + \! \frac{1}{3} \metric_{\rho \sigma} \metric^{\mu [\alpha} \delta_{\lambda}^{\beta]} \right] \! \delta \Delta t^{\rho \lambda \sigma} \text{.}
\end{align}
Hence, the variations of the action yield
\begin{align}
 \frac{\delta S_{\Phi}}{\delta \Delta v^{\mu \nu}} ={}& \frac{1}{4\pi G} \big( (\Delta v_{\kappa}^{\point \kappa}) g_{\mu \nu} - \Delta v_{\mu \nu} \big) \text{,}\\ 
 \frac{\delta S_{\Phi}}{\delta \Delta a^{\mu \nu}} ={}& \frac{1}{4\pi G} \Delta a_{\mu \nu} \text{,} \\
 \frac{\delta S_{\Phi}}{\delta \Delta t^{\mu \alpha \beta}} ={}& \frac{4}{\pi G} \Delta t_{[\mu \alpha] \beta} \text{.}
\end{align}
Imposing an action principle $\delta S_{\Phi} = 0$ this requires the
spin torsion to vanish in the absence of sources or boundary
conditions for this simplest classical theory. The resulting theory is
identical to classical general relativity.

We conclude this section with two additional remarks: First, a
different definition of spin curvature would be suggested in the
presence of Rarita-Schwinger spinors $\psi^{\lambda}$, being a first-rank tensor
in spacetime as well as in Dirac space. In analogy to \Eqref{eq:Riemann}, we would define
\begin{align}
 \Phi_{\mu \nu \point \rho}^{\point[2] \lambda} \psi^{\rho} = [ \nabla_{\mu} , \nabla_{\nu} ] \psi^{\lambda} + C^{\sigma}_{\point \mu \nu} \nabla_{\sigma} \psi^{\lambda}.
\end{align}
This spin curvature can be decomposed into
\begin{align}
 \Phi_{\mu \nu \lambda \rho} = R_{\mu \nu \lambda \rho} \, \mrI + \Phi_{\mu \nu} \, \metric_{\lambda \rho} \text{.}
\end{align}
Here, the spacetime curvature tensor $R_{\mu \nu \lambda \rho}$
appears as the antisymmetric part of $\Phi_{\mu \nu \lambda \rho}$ in
$\lambda \leftrightarrow \rho$ and the previous spin curvature
$\Phi_{\mu \nu} \metric_{\lambda \rho}$ arises as the symmetric
term. Forming suitable first order invariants of this spin curvature,
we end up with actions of Einstein-Hilbert type including both
spacetime and spin curvature. In the spirit of the speculative
interpretation given above, the absence of spacetime torsion in our
universe would fit well to a non-existence of fundamental
Rarita-Schwinger fields.

For our second remark, we disregard any torsion such that
$\Phi_{\mu\nu} \to \cplx \mcF_{\mu \nu} + \hat{\Phi}_{\mu\nu}$. In this case, the second-order
invariant of the spin curvature which is reminiscent to the kinetic
term of a gauge theory reduces to
\begin{equation}
\frac{1}{d_{\mcA} d_{\gamma}}\tr \Phi_{\mu\nu}\Phi^{\mu\nu} \to - \frac{1}{d_{\mcA}} \tr \mcF_{\mu\nu} \mcF^{\mu\nu} - \frac{1}{8} R_{\mu \nu \rho \lambda} R^{\mu \nu \rho \lambda} \text{.}
\end{equation}
Naively, this seems to suggest that a gauge-gravity field theory links
the coupling to the gauge fields to that of higher-order curvature
terms. However, this connection can, of course, simply be broken
explicitly by additional $\mcF_{\mu\nu} \mcF^{\mu\nu}$ terms in the
action which are not part of a $\Phi_{\mu\nu}\Phi^{\mu\nu}$ term.

\section{Reducible Representations}
\label{sec:red_rep}

So far, our considerations have been based onto the irreducible
representation of the Clifford algebra characterized by $d_\gamma=4$
in four spacetime dimensions. A generalization of our formalism to
reducible representations is not completely trivial, since the
construction of the spin connection makes explicit use of a particular
complete basis of the Clifford algebra. The basis used above may not
generalize straightforwardly to any reducible
representation. Therefore, we confine ourselves to those reducible
representations where the basis used so far is still sufficient.

Our construction leads to reducible representations with $d_{\gamma} =
4 n$, for $n \in \N$. For this, we assume that the new Dirac matrices
can be written as tensor product of a possibly spacetime dependent
matrix $A \in \C^{n \times n}$ of dimension $n$ and the Dirac matrices
$\gamma^{\mu} \in \C^{4 \times 4}$ of the irreducible representation
used above,
\begin{align}
 {\gamma_\ttm{(d_{\gamma})}}^{\mu} = A \otimes \gamma^{\mu},
\end{align}
obviously implying that $d_\gamma=4n$. Of course, the set of
${\gamma_\ttm{(d_{\gamma})}}^{\mu}$ shall also satisfy the Clifford
algebra
\begin{align}
 \{ {\gamma_\ttm{(d_{\gamma})}}^{\mu} , {\gamma_\ttm{(d_{\gamma})}}^{\nu} \} = 2 g^{\mu \nu} \mrI_\ttm{(d_{\gamma})} = 2 g^{\mu\nu} \mrI_\ttm{(n)} \otimes \mrI
\end{align}
which tells us that $A$ is idempotent.
\begin{align}
 A^2 = \mrI_\ttm{(n)}
\end{align}
has to hold at any spacetime point. Analogous to our previous
construction, we need a covariant derivative
${\nabla_\ttm{(d_{\gamma})}}_{\mu}$ and a spin metric
$h_\ttm{(d_{\gamma})}$. We require the covariant derivative to
factorize accordingly,
\begin{align}
 {\nabla_\ttm{(d_{\gamma})}}_{\mu} A \otimes \gamma^{\nu} ={}& ( {\nabla_\ttm{(n)}}_{\mu} A ) \otimes \gamma^{\nu} + A \otimes (\nabla_{\mu} \gamma^{\nu}),
\end{align}
where ${\nabla_\ttm{(n)}}_{\mu}$ acts on the `$A$-part' and
$\nabla_{\mu}$ is identical to the covariant derivative in irreducible
representation. Analogously to \Eqref{eq:cov_deriv_gamma}, we also
demand for
\begin{align}
 {\nabla_\ttm{(d_{\gamma})}}_{\mu} A \otimes \gamma^{\nu} ={}& D_{\mu} A \otimes \gamma^{\nu} + [ {\Gamma_\ttm{(d_{\gamma})}}_{\mu} , A \otimes \gamma^{\nu} ],
\end{align}
which tells us that the affine connection has to read
\begin{align}
 {\Gamma_\ttm{(d_{\gamma})}}_{\mu} = {\Gamma_\ttm{(n)}}_{\mu} \otimes \mrI + \mrI_\ttm{(n)} \otimes \Gamma_{\mu} \text{.}
\end{align}
Because the irreducible component already carries all relevant
structures for general covariance, the $A$-part in its simplest form
should be covariantly constant,
\begin{align}
 0 = {\nabla_\ttm{(n)}}_{\mu} A = \partial_{\mu} A + [ {\Gamma_\ttm{(n)}}_{\mu} , A ] \text{.}
\end{align}
We can rewrite this into a condition for the connection
${\Gamma_\ttm{(n)}}_{\mu}$ which has to satisfy
\begin{align}
  {\Gamma_\ttm{(n)}}_{\mu} = A  {\Gamma_\ttm{(n)}}_{\mu} A^{-1} - (\partial_{\mu} A) A^{-1} \text{.}
\label{eq:defGamman}
\end{align}
For a given choice of $A$ on a given spacetime, \Eqref{eq:defGamman}
may or may not have a solution in terms of a set of
${\Gamma_\ttm{(n)}}_{\mu}$. If a solution exists, it completes the
definition of the spin connection for this reducible representation.
For the simpler case of constant matrices $A$, a solution is always
given by ${\Gamma_\ttm{(d_{\gamma})}}_{\mu} = 0$.

The natural way to embed the spin-base transformations is given by the form
\begin{align}
 \mcS_\ttm{(d_{\gamma})} = \mrI_\ttm{(n)} \otimes \mcS \text{,} \quad \mcS \in \text{SL}(4, \C) \text{.}
\label{eq:embed}
\end{align}
The corresponding transformation law for the spin connection then reads
\begin{align}
 {\Gamma_\ttm{(d_{\gamma})}}_{\mu} \rightarrow \mcS_\ttm{(d_{\gamma})} {}& {\Gamma_\ttm{(d_{\gamma})}}_{\mu} {\mcS_\ttm{(d_{\gamma})}}^{-1} - (\partial_{\mu} \mcS_\ttm{(d_{\gamma})}) {\mcS_\ttm{(d_{\gamma})}}^{-1} \notag\\
 ={}& {\Gamma_\ttm{(d_{\gamma})}}_{\mu} \otimes \mrI + \mrI_\ttm{(n)} \otimes (\mcS \Gamma_{\mu} \mcS^{-1} - (\partial_{\mu} \mcS) \mcS^{-1}) \text{.}
\end{align}
It is worthwhile to emphasize that the choice of the embedding
\eqref{eq:embed} is not unique. Reducible representations of the
Clifford algebra have a much larger symmetry of SL$(d_\gamma=4n,\C)$,
such that there are typically many more options of embedding
SL($4,\C$) into SL$(d_\gamma=4n,\C)$. The present choice is motivated
by the similarity to the embedding of local Lorentz transformations
that we would encounter in the corresponding vierbein formalism.
Vierbeins transform under these Lorentz transformations as
\begin{align}
 \vierbein^{\mu}_{\point a} \rightarrow \vierbein^{\mu}_{\point b} \Lambda^{b}_{\point a} \text{,}
\end{align}
corresponding on the level of Dirac matrices to
\begin{align}
 {\gamma_\ttm{(\vierbein)}}^{\mu} \rightarrow \mcS_{\mathrm{Lor}} {\gamma_\ttm{(\vierbein)}}^{\mu} {\mcS_{\mathrm{Lor}}}^{-1} \text{.}
\end{align}
The matrix $\mcS_{\mathrm{Lor}}$ is given by
\begin{align}
 \mcS_{\mathrm{Lor}} = \exp \! \left( \frac{\eta_{a c} \omega^{c}_{\point b}}{8} [{\gamma_\ttm{(\mathrm{f})}}^{a} , {\gamma_\ttm{(\mathrm{f})}}^{b}] \right) \! \text{,}
\end{align}
where the matrix $(\omega^{a}_{\point b})$ is defined by
\begin{align}\
 \Lambda^{a}_{\point b} = (\euler^{\omega})^{a}_{\point b} \text{.}
\end{align}
Promoting the (fixed) Dirac matrices to the reducible representation given above, the corresponding Lorentz transformation reads
\begin{align}
 \mcS_{\mathrm{Lor} \ttm{(d_{\gamma})}} = \exp \! \left( \frac{\eta_{a c} \omega^{c}_{\point b}}{8} [ A \otimes {\gamma_\ttm{(\mathrm{f})}}^{a}, A \otimes {\gamma_\ttm{(\mathrm{f})}}^{b} ] \right) \! \equiv \mrI_{n} \otimes \mcS_{\mathrm{Lor}} \text{,}
\end{align}
which is structurally identical to our choice for the embedding of \Eqref{eq:embed}.

Finally, we also need the spin metric for the reducible
representation, which has to satisfy
\begin{align}
 {\gamma_\ttm{(d_{\gamma})}}_{\mu}^{\dagger} = - h_\ttm{(d_{\gamma})} {\gamma_\ttm{(d_{\gamma})}}_{\mu} {h_\ttm{(d_{\gamma})}}^{-1} \text{.}
\end{align}
It is obvious that this condition is satisfied by
\begin{align}
 h_\ttm{(d_{\gamma})} = A \otimes h \text{,} \quad A^{\dagger} = A,
\label{eq:embedh}
\end{align}
demanding that $A$ is hermitean in order to have
$h_\ttm{(d_{\gamma})}$ antihermitean. Of course also the absolute
value of the determinant is equal to one as required, since
\begin{align}
 \abs{\det h_\ttm{(d_{\gamma})}} = \sqrt{ \abs{ \det A \otimes h}^2} = \sqrt{ \abs{ \det \mrI_\ttm{(n)} \otimes h^2 }} = 1 \text{.}
\end{align}
This completes the construction of a generalization to particularly
simple reducible representations of the Dirac algebra.

Again, the embedding \eqref{eq:embedh} may not be unique. The
present choice is intuitive, because in conventional choices for
the flat spacetime Dirac matrices, the spin metric is simply given by
${\gamma_\ttm{(\mathrm{f})}}_{0}$. In the corresponding reducible representation, the `new'
${\gamma_\ttm{(d_{\gamma)}}}_{0}$ would read
\begin{align}
 {\gamma_\ttm{(d_{\gamma)}}}_{0} = A \otimes {\gamma_\ttm{(\mathrm{f})}}_{0} \text{,}
\end{align}
matching precisely with our extended spin metric.

Let us emphasize again that the straightforwardly induced
symmetries of the present construction may not exhaust the full
invariance of the reducible Clifford algebra. For instance, one can
immediately verify that our construction is invariant under local
SU($n$)$\otimes$SL($4,\C$) transformations, which is
in general only a subgroup of the SL($d_\gamma,\C$) invariance of
the Clifford algebra in reducible representation.

\section{Path integral}
\label{sec:pathintegral}

As an application of the spin-base invariant formalism, let us discuss
possible implications for quantizing gravity within a path integral
framework. Of course, the question as to whether such a path integral
exists is far from being settled. For the purpose of the following
discussion, we simply assume that there is such a path integral
possibly regularized in a symmetry-preserving way and possibly
amended with a suitable gauge fixing procedure. For simplicity, we
consider the case of vanishing spin torsion, spacetime torsion and
gauge fields
\begin{align}
 \Delta \Gamma_{\mu} = 0 \text{,} \quad C^{\kappa}_{\point \mu \nu} = 0 \text{,} \quad \mcA_{\mu} = 0 \text{,}
\end{align}
even though the following considerations will not interfer with any of
these quantities. Also, we work manifestly in $d = 4$ where
$d_{\gamma} = 4$.

So far, we took the viewpoint that the spacetime-dependent Dirac
matrices $\gamma_\mu$ are the basic objects encoding the essential
properties of the spacetime. In fact, given a set of Dirac matrices,
we can compute the metric,
\begin{align}
 g_{\mu \nu} = \frac{1}{{4}} \tr(\gamma_{\mu} \gamma_{\nu}) \text{.}
\end{align}
Also the spin metric necessary for including Fermionic Dirac degrees
of freedom is fixed (up to a sign) by the condition
\begin{align}
 h^{\dagger} = - h \text{,} \quad \bar{\gamma}^{\mu} = - \gamma^{\mu} \text{,} \quad \abs{\det h} = 1 \text{,}
\end{align}
see App. \ref{App:spin_metric} and
Eqs. \eqref{eq:spin_metric}-\eqref{eq:spin_metric2}. The Dirac
matrices also determine the spin connection (up to spin torsion),
cf. \Eqref{eq:spintorsionconstraints}, and all these ingredients
suffice to define a classical theory of gravity including dynamical
fermions. One is hence tempted to base a quantized theory also on the
Dirac matrices as the fundamental degree of freedom. This would be
analogous to quantizing gravity in terms of a vierbein. Whereas this
is certainly a valid and promising option, we show in the following
that this Dirac matrix/vierbein quantization is actually not
necessary.

Demanding that quantization preserves the local Clifford algebra
constraint also off shell
\begin{align}
 \{ \gamma^{\mu} , \gamma^{\nu} \} = 2 \metric^{\mu \nu} \mrI , \quad \gamma^{\mu} \in \C^{{4} \times {4}},
\label{def:Clifford3}
\end{align}
(for a correspondingly off-shell metric), the Weldon theorem
\eqref{eq:variation_gamma} already tells us that a fluctuation of the
Dirac matrices can always be decomposed into a metric fluctuation and
an SL$(4,\C)_\gamma$ fluctuation,
\begin{align}
 \delta \gamma^{\mu} = \frac{1}{2} (\delta \metric^{\mu \nu}) \gamma_{\nu} + [\delta \mcS_{\gamma} , \gamma^{\mu}] \text{.} \label{eq:variation_gamma2}
\end{align}
Hence, we do not attempt to construct an integration measure for
Dirac matrices ``$\mathcal{D} \gamma$'', satisfying the Dirac algebra
constraint. Instead, it appears more natural to integrate over
metrics and SL$(4,\C)_\gamma$ fluctuations. In the following,
we show that the SL$(4,\C)_\gamma$ fluctuations factor out of
the path integral because of spin-base invariance, such that a
purely metric-based quantization scheme appears sufficient also in
the presence of dynamical fermions.

The crucial starting point of our line of argument is the fact that
all possible sets of Dirac matrices compatible with a given
metric are connected with each other via SL$(4,\C)_\gamma$
transformations \cite{Schmutzer:1968}. This means that we can cover
the space of Dirac matrices by (i) choosing an arbitrary mapping
$\tilde{\gamma}^{\mu}$ of the metric into the space of Dirac matrices
satisfying the Clifford algebra
\begin{align}
 \metric_{\mu \nu} \rightarrow \tilde{\gamma}_{\mu} = \tilde{\gamma}_{\mu}(\metric),
\end{align}
and (ii) performing SL$(4,\C)_\gamma$ transformations
$\euler^{\mcM}$ of this mapping
\begin{align}
 \gamma_{\mu}(\metric) = \gamma_{\mu}\big(\tilde{\gamma}(\metric) , \mcM(\metric) \big) = \euler^{\mcM(\metric)} \tilde{\gamma}_{\mu}(\metric) \euler^{-\mcM(\metric)} \label{eq:Diracmapping2}
\end{align}
where $\mcM$ is an arbitrary tracefree matrix which can be spanned by
the generators of SL$(4,\C)_\gamma$ transformations. This matrix $\mcM$
may even depend on the metric if we demand 
$\gamma_{\mu}(\metric)$ to be a particular Dirac matrix compatible with
the Clifford algebra independently of the choice of the
representative Dirac matrix $\tilde{\gamma}_\mu$.

Equation \eqref{eq:Diracmapping2} emphasizes the fact that
every possible set of Dirac matrices yielding a given metric
$\metric_{\mu\nu}$ can be constructed by this mapping.

The variation of the resulting Dirac matrices under an infinitesimal
variation in terms of the metric $\delta \metric_{\mu \nu}$ can be
represented analogously to the Weldon theorem:
\begin{align}\label{eq:var_gamma}
 \delta \gamma_{\mu} ={}& \frac{1}{2} (\delta \metric_{\mu \nu}) \gamma^{\nu} + \left[ G^{\rho \lambda} \, \delta \metric_{\rho \lambda} , \gamma_{\mu}\right],
\end{align}
where the tensor $G_{\rho \lambda}$ is tracefree and depends on the
actual choice of $\tilde{\gamma}_{\mu}(\metric)$ and
$\mcM(\metric)$. $G^{\rho \lambda}$ can be calculated from
\begin{align}
 \big[ [ G^{\rho \lambda} , \gamma_{\mu} ] , \gamma^{\mu} \big] = \left[ \frac{\partial \gamma_{\mu}(\metric)}{\partial \metric_{\rho \lambda}} , \gamma^{\mu} \right] \text{.}
\end{align}
The infinitesimal SL$(4,\C)_\gamma$ fluctuation $\delta \mcS_{\gamma}$
acting on the Dirac matrices, as it occurs in the Weldon theorem, is
obviously given by
\begin{align}
 \delta \mcS_{\gamma} = G^{\rho \lambda} \, \delta \metric_{\rho \lambda} \text{.}
\end{align}
Now, the microscopic actions subject to quantization are considered to
be functionals of the fermions and the Dirac matrices,
$S[\psi,\bar{\psi},\gamma]$. From our construction given above, the
Dirac matrices arise from a representative Dirac matrix
$\tilde{\gamma}^{\mu}(\metric)$ which is related to the metric by an
arbitrary but fixed \textit{bijection}, $\metric_{\mu \nu}
\leftrightarrow \tilde{\gamma}^{\mu}$. The Dirac matrix $\gamma^{\mu}$
occuring in the action is then obtained via the
SL$(4,\C)_\gamma$ transformation governed by $\mcM$,
cf. \Eqref{eq:Diracmapping2}. Therefore, it is useful to think of the
action as a functional of the metric and of $\mcM$,
$S[\psi,\bar{\psi},\metric; \mcM]$. In particular, the freedom to
choose $\mcM$ (or the corresponding SL$(4,\C)_\gamma$ group
element) guarantees that the space of all possible Dirac matrices
compatible with a given metric can be covered -- for any choice of the
representative $\tilde{\gamma}^{\mu}(\metric)$.

In addition to diffeomorphism invariance, we demand that the actions
under consideration are invariant under spin-base transformations
\begin{align}
 S[\psi,\bar{\psi},\metric; \mcM] \! \rightarrow \! S[\mcS \psi , \bar{\psi} \mcS^{-1} , \metric; \ln (\mcS \euler^{\mcM})] \! \equiv \! S[\psi, \bar{\psi},\metric ; \mcM] \text{.}
\end{align}
Especially we may always choose
\begin{align}
 \mcS = \euler^{- \mcM},
\end{align}
such that
\begin{align}
 S[\psi,\bar{\psi},\metric; \mcM] = S[\psi',\bar{\psi}',\metric;0] \text{,} \,\,\, \psi' = \euler^{- \mcM} \psi  \text{,} \,\,\, \bar{\psi}' = \bar{\psi} \euler^{\mcM} \text{.}
\end{align}
The essential ingredient for a path integral quantization is the
choice of the measure. As argued above, the present construction
suggests, to integrate over metrics $g$ and successively over
$\mcM$ to cover the space of all Dirac matrices.

More specifically, let us study the expectation value of an operator
$\hat{\Obs}(\psi,\bar{\psi},\metric;\mcM)$ which is a scalar under
spin-base transformations. For illustrative purposes, let us first consider
only the functional integrations over the fermion and metric degrees of freedom:
\begin{align}
 {\Obs}[\mcM] ={}& \langle \hat\Obs(\psi,\bar{\psi},\metric; \mcM) \rangle\\
 ={}& \int \!\!\! \mcD \metric \mcD \psi \mcD \bar{\psi} \, \hat\Obs(\psi,\bar{\psi},\metric;\mcM) \, \euler^{\cplx S[\psi,\bar{\psi},\metric;\mcM]} \text{,}
\end{align}
with suitable measures $\mcD \metric \mcD \psi \mcD \bar{\psi}$. The
following argument only requires that the measure transforms
in a standard manner under a \textit{change of variables}
\begin{align}
 \mcD \psi = \mcD \psi' \left( \det \frac{\delta \psi'}{\delta \psi} \right)^{-1}.
\end{align}
As a consequence, $\mcD \psi \mcD \bar{\psi}$ is invariant under
spin-base transformations, since the Jacobians from $\mcD
\psi$ and from $\mcD \bar{\psi}$ are inverse to each other
\begin{align}
 \mcD \psi \mcD \bar{\psi} = \mcD (\mcS \psi) \mcD (\bar{\psi} \mcS^{-1}) \text{.}
\end{align}
Because $\hat\Obs$ is a scalar in Dirac space, it also needs to be
invariant under spin-base transformations
\begin{align}
\begin{aligned}
 \hat\Obs(\psi,\bar{\psi},\metric; \mcM) \rightarrow \hat\Obs {}&\big(\mcS \psi,\bar{\psi} \mcS^{-1},\metric; \ln( \mcS \euler^{\mcM})\big)\\
 {}& \equiv \hat\Obs(\psi,\bar{\psi},\metric; \mcM) \text{.}
\end{aligned}
\end{align}
Now it is easy to see, that ${\Obs}[\mcM]$ is actually independent of
the choice of $\mcM(\metric)$
\begin{align}
{\Obs}[\mcM] ={}& \int \!\!\! \mcD \metric \mcD \psi \mcD \bar{\psi} \, \hat\Obs(\psi,\bar{\psi},\metric;\mcM) \, \euler^{\cplx S[\psi,\bar{\psi},\metric;\mcM]} \notag\\
 ={}& \int \!\!\! \mcD \metric \mcD \psi' \mcD \bar{\psi}' \, \hat\Obs(\psi',\bar{\psi}',\metric;0) \, \euler^{\cplx S[\psi',\bar{\psi}',\metric;0]} \notag\\
 ={}& {\Obs}[0] \text{.}
\end{align}
Therefore, every set of Dirac matrices compatible with a given metric
contributes indentically to such an expectation value. Hence, we may
choose any convenient spin basis to simplify explicit
computations. From another viewpoint, an additional functional
integration over SL$(4, \C)_\gamma$ with a suitable measure
$\mcD \mcM$ would have factored out of the path integral and thus can
be included trivially in its normalization.

This concludes our argument that a quantization of interacting
theories of fermions and gravity may be solely based on a quantization
of the metric together with the fermions. The spin-base invariant
formulation given here suggests that this quantization scheme is
natural. A quantization in terms of vierbeins/Dirac matrices -- though
perhaps legitimate -- is not mandatory.

With hindsight, our results rely crucially on the constraint that the
fluctuations of the Dirac matrices satisfy the Clifford algebra
\Eqref{def:Clifford3} also off-shell. If this assumption is
relaxed, e.g., if the anticommutator of two Dirac matrices in the path
integral is no longer bound to be proportional to the identity, a
purely metric-based quantization scheme may no longer be possible.

\section{Metric variations in the spin-base invariant formalism}
\label{sec:variations}

In this section, we discuss the response of several objects under
variations of the metric, yielding a set of properties that may become
relevant in concrete quantum gravity computations within the present
formalism. The formalism has already been used successfully for
theories with quantized fermions in curved spacetime
\cite{Casals:2012es,Gies:2013dca}.

For both perturbative as well as non-perturbative calculations,
propagators are central objects. As they arise from two-point
correlators, we study the response of several quantities up to second
order in metric fluctuations in the following. Since field theory
calculations generically need a spacetime ``to stand on'', we
introduce a fiducial background metric $\bar\metric$ with respect to
which variations are performed. 

Let us first consider the variation of the Dirac matrices to second
order in the fluctuations around this background,
\begin{align}\label{eq:gamma_metric_expansion}
\begin{aligned}
 \gamma_{\mu}(\bar{\metric}+\delta\metric) ={}& \bar{\gamma}_{\mu} + \left.\frac{\partial \gamma_{\mu}(\metric)}{\partial \metric_{\rho \lambda}}\right|_{\metric = \bar{\metric}} \delta \metric_{\rho \lambda} \\
 {}&+ \frac{1}{2} \left. \frac{\partial^2 \gamma^{\mu}(\metric)}{\partial \metric_{\alpha \beta} \partial \metric_{\rho \lambda}} \right|_{\metric = \bar{\metric}} \delta \metric_{\alpha \beta} \delta \metric_{\rho \lambda} + \mcO(\delta \metric^{3}) \text{,}
\end{aligned}
\end{align}
where $\bar{\gamma}_{\mu} =
\gamma_{\mu}(\bar{\metric})$.\footnote{Within the present
section, the bar only refers to the background-fields quantities and not the Dirac
conjugation; $\bar{\gamma}_{\mu}$ here should thus not be
confused with $- h \gamma_{\mu}^{\dagger} h^{-1}$.} From
\Eqref{eq:var_gamma} we get
\begin{align}\label{eq:first_derivative_gamma}
 \frac{\partial \gamma_{\mu} (\metric)}{\partial \metric_{\rho \lambda}} = \frac{1}{2} \delta_{\mu \nu}^{\rho \lambda} \, \gamma^{\nu}(\metric) + \big[ G^{\rho \lambda} ( \metric ) , \gamma_{\mu} ( \metric ) \big] \text{,}
\end{align}
with the \textit{symmetrized} product of two deltas, $\delta^{\rho
  \lambda}_{\mu \nu} = \frac{1}{2}(\delta^{\rho}_{\mu}
\delta^{\lambda}_{\nu} + \delta^{\rho}_{\nu}
\delta^{\lambda}_{\mu})$. We know that the first part on the
right-hand side is obligatory and therefore cannot be eliminated by
spin-base transformations. But the second term is only a variation of
the spin base and can thus be transformed to zero, at least for the background field metric. Therefore we may demand
\begin{align}\label{eq:optimal_choice_G_1}
 G^{\rho \lambda}(\bar{\metric}) = 0 \text{,}
\end{align}
which corresponds to implicitly choosing part of the spin base.

Assuming that $\gamma_{\mu}(\metric)$ is a sufficiently
smooth function of the metric, partial derivatives with respect to
different metric components commute. This constrains the first
derivative of $G^{\rho \lambda}(\metric)$ which we call
$G^{\alpha \beta \rho \lambda} = \left. \frac{\partial G^{\rho
    \lambda} (\metric)}{\partial \metric_{\alpha
    \beta}} \right|_{\metric = \bar{\metric}}$.
It is useful to introduce the auxiliary tensor
\begin{align}
 \omega^{\rho \lambda \alpha \beta}_{\point[4] \mu \nu} = \frac{1}{4} \delta_{\mu \kappa}^{\rho \lambda} \, \bar{\metric}^{\kappa \sigma} \delta_{\sigma \nu}^{\alpha \beta} = \omega^{\alpha \beta \rho \lambda}_{\point[4] \nu \mu} \text{,}
\end{align}
which shows up in
\begin{align}\label{eq:second_derivative_gamma}
 {}& \left. \!\!\!\!\! \frac{\partial^2 \gamma_{\mu}(\metric)}{\partial \metric_{\alpha \beta} \partial \metric_{\rho \lambda}} \right|_{\metric = \bar{\metric}} = - \omega^{\rho \lambda \alpha \beta}_{\point[4] \mu \nu} \bar{\gamma}^{\nu} + [G^{\alpha \beta \rho \lambda}, \bar{\gamma}_{\mu}] \notag\\
 {}& = - \omega^{\alpha \beta \rho \lambda}_{\point[4] (\mu \nu)} \bar{\gamma}^{\nu} + \left[ G^{\alpha \beta \rho \lambda} - \frac{1}{8} \omega^{\alpha \beta \rho \lambda}_{\point[4] [\kappa \sigma]} [ \bar{\gamma}^{\kappa} , \bar{\gamma}^{\sigma}] , \bar{\gamma}_{\mu} \right] \! \text{.}
\end{align}
Again the first part is obligatory since it is symmetric in $(\alpha
\beta) \leftrightarrow (\rho \lambda)$ and arises from the first term
of \Eqref{eq:first_derivative_gamma}. Therefore the simplest choice is
\begin{align}\label{eq:optimal_choice_G_2}
 G^{\alpha \beta \rho \lambda} = \frac{1}{8} \omega^{\alpha \beta \rho \lambda}_{\point[4] [\kappa \sigma]} [ \bar{\gamma}^{\kappa} , \bar{\gamma}^{\sigma}] \text{,}
\end{align}
leading to the simple variation to second order in $\delta\metric$,
\begin{align}\label{eq:simplest_gamma_result}
 \gamma_{\mu}(\bar{\metric} + \delta \metric) \simeq \bar{\gamma}_{\mu} \! + \! \frac{1}{2} \delta^{\rho \lambda}_{\mu \nu} \bar{\gamma}^{\nu} \delta \metric_{\rho \lambda} \! - \! \frac{1}{2} \omega^{\alpha \beta \rho \lambda}_{\point[4] (\mu \nu)} \bar{\gamma}^{\nu} \delta \metric_{\alpha \beta} \delta \metric_{\rho \lambda} \text{.}
\end{align}
Of course, if different conditions on $G^{\rho\lambda}$ or $G^{\alpha
  \beta \rho \lambda}$ are imposed, the variation of the Dirac
matrices will have a different form.

With this result (or corresponding results for other conditions on
$G^{\rho\lambda}$ or $G^{\alpha \beta \rho \lambda}$), variations of
field monomials formulated in terms of the Dirac matrices with respect
to the metric can be calculated straightforwardly. Immediate
applications are the computation of the Hessian of a bare action,
corresponding to the inverse bare graviton propagator, or a Hessian of
an effective action, yielding the full propagator.

If further dynamical fermion fields are included, we also need the
variations of the spin metric, etc., at least in principle. In
practice, they turn out to be irrelevant at the two-point level, as
demonstrated now: For example, the variation of the spin metric has to
satisfy \Eqref{eq:gamma_dirac_conj},
\begin{align}
 (\bar{\gamma}_{\mu} + \delta \gamma_{\mu})^{\dagger} ={}& - ( \bar{h}
  + \delta h) (\bar{\gamma}_{\mu} + \delta \gamma_{\mu}) ( \bar{h} +
  \delta h )^{-1} \text{,}
\end{align}
where $\bar{h}$ is the spin metric corresponding to
$\bar{\gamma}_{\mu}$. The variations $\delta h$ and $\delta
\gamma_{\mu}$ parametrize the deviations of $h(\bar{\metric} + \delta
  \metric)$ and $\gamma_{\mu}(\bar{\metric} + \delta
  \metric)$ from the background-field quantities. For
our choice \Eqref{eq:simplest_gamma_result}, we have $ \delta \gamma_{\mu}^{\dagger} = - \bar{h} (\delta
\gamma_{\mu}) \bar{h}^{-1}$ neglecting terms with $\mcO(\delta
\metric^3)$.  This equation leads to
\begin{align}\label{eq:variation_spin_metric}
 0 \simeq [(\bar{\gamma}_{\mu} + \delta \gamma_{\mu})(\mrI -
   \bar{h}^{-1} \delta h) , \bar{h}^{-1} \delta h] \text{.}
\end{align}
Here we have used, that $\delta h$ is at least of order $\delta \metric$ such that
we only need to keep track of all terms up to order $\delta \metric$
within the other terms, yielding
\begin{align}
\begin{aligned}
 0 \simeq{}& [ \bar{\gamma}_{\mu} + \delta \gamma_{\mu} , \bar{h}^{-1}
   \delta h] \\ \simeq{}& \left(\bar{\metric}_{\mu \rho} + \frac{1}{2}
 \delta \metric_{\mu \rho} \right) [ \bar{\gamma}^{\rho} ,
   \bar{h}^{-1} \delta h]
\end{aligned}
\end{align}
by multiplying \Eqref{eq:variation_spin_metric} from the right with
$\mrI + \bar{h}^{-1} \delta h$. Multiplying by
$\bar{\metric}^{\nu \mu} - \frac{1}{2} \bar{\metric}^{\nu \alpha}
(\delta \metric_{\alpha \beta}) \bar{\metric}^{\beta \mu}$, we find
\begin{align}
 \delta h = \varepsilon \bar{h} + \mcO(\delta \metric^3) \text{,}
 \quad \varepsilon \in \R,
\end{align}
for an arbitrary infinitesimal $\varepsilon$, which needs to be real
because $\delta h$ needs to be antihermitean. But $h(\bar{\metric} +
  \delta \metric)$ still needs to have a determinant
with absolute value equal to one, resulting in a constraint for
$\varepsilon$
\begin{align}
 1 = \abs{ \det (\bar{h} + \delta h)} = \abs{ \det( \bar{h} \big( \mrI
   + \bar{h}^{-1} \delta h) \big) } \simeq (1 + \varepsilon)^4
 \text{.}
\end{align}
This equation only has two real solutions $\varepsilon_{1} = -2$ and
$\varepsilon_{2} = 0$. Of course, $\varepsilon_{1}$ is not
infinitesimal but corresponds to the discrete transformation $\bar{h}
\rightarrow - \bar{h}$. This solution reflects the ambiguity in the
choice of the sign of the spin metric and therefore is irrelevant. The
relevant second solution shows that the spin metric is constant to
second order in the metric variation
\begin{align}
 h(\bar{\metric} + \delta \metric) = \bar{h} +
 \mcO(\delta \metric^{3}) \text{.}
\end{align}
Analogously, it can be derived from $\{ \gamma_{\ast} , \gamma_{\mu}
\} = 0$ and $\gamma_{\ast}^2 = \mrI$ that also $\gamma_{\ast}$ is
constant to second order
\begin{align}
 \gamma_{\ast}(\bar{\metric} + \delta \metric) = \gamma_{\ast}(\bar{\metric}) + \mcO (\delta \metric^3) \text{.}
\end{align}
Finally, let us study the variation of the spin connection
$\Gamma_{\mu}$. For the spin torsion $\Delta \Gamma_{\mu}$ this is
particularly simple, as it depends on the metric only through the base
elements $\gamma_{\mu}, \gamma_{\ast} \gamma_{\mu} , [ \gamma_{\mu} ,
  \gamma_{\nu}]$ the variations of which are straightforward.  The
variation of the connection $\hat{\Gamma}_{\mu}$ can also be
straightforwardly worked out using ${D_\ttm{(\mathrm{LC})}}_{\mu}
\gamma_{\nu} = - [ \hat{\Gamma}_{\mu} , \gamma_{\nu}]$ and $\tr
\hat{\Gamma}_{\mu} = 0$. We find
\begin{align}
 {}&\hat{\Gamma}_{\mu}(\bar{\metric} + \delta \metric) = \hat{\Gamma}_{\mu}(\bar{\metric}) + \delta \hat{\Gamma}_{\mu} + \mcO(\delta \metric^3)\\
{}&\begin{aligned}
 \delta \hat{\Gamma}_{\mu} ={}& \frac{1}{8} [ \bar{\gamma}^{\kappa} , \bar{\gamma}^{\sigma}] \! \left[ \delta^{\alpha \beta}_{\mu [ \kappa} \delta^{\nu}_{\sigma]} + \delta \metric_{\rho \lambda} \left( \vphantom{\frac{1}{2}} \omega^{\alpha \beta \rho \lambda}_{\point[4] [\kappa \sigma]} \delta_{\mu}^{\nu} \right. \right. \\
 {}& \hspace{0.5cm} \left. \left. - 2 \omega^{\alpha \beta \rho \lambda}_{\point[4] \mu [ \kappa} \delta^{\nu}_{\sigma]} - \frac{1}{2} \delta^{\alpha \beta}_{\mu [\kappa} \delta^{\rho \lambda}_{\sigma] \chi} \bar{\metric}^{\chi \nu} \right) \right] \! \bar{D}_\ttm{(\mathrm{LC})} {{}_{}}_{\nu} \delta \metric_{\alpha \beta} \text{.}
\end{aligned}
\end{align}

In a certain sense, our conditions on $G^{\rho\lambda}$ or $G^{\alpha
  \beta \rho \lambda}$ represent a minimal choice as they minimize the
number of terms present in the variation of the Dirac matrices to the
corresponding order. Calculations should therefore simplify in
comparison to other choices. Due to the direct relation of
$G^{\rho\lambda}$ to spin base transformations, it is obvious that
physical observables are independent of the choice of conditions.

It is interesting to note, that the variation of the Dirac matrices
\Eqref{eq:gamma_metric_expansion} using the conditions
\Eqref{eq:optimal_choice_G_1} and (\ref{eq:optimal_choice_G_2})
corresponds exactly to the result obtained within the vierbein
formalism if the Lorentz symmetric gauge is used
\cite{Woodard:1984sj,vanNieuwenhuizen:1981uf}. This gauge has already
proved to be a useful choice within the vierbein formalism, and
has for instance been used in a functional RG calculation of
fermions in quantized gravity in \cite{Eichhorn:2011pc}. Hence, our
choice can be viewed as the direct generalization of the Lorentz
symmetric gauge into the spin-base invariant formulation.

Let us finally comment on the differences between our
metric-based quantization scheme and vierbein- (or Dirac
matrix-)based schemes both of which are a priori legitimate
strategies for quantization. An obvious difference occurs in the
corresponding Hessians: given a bare or effective action $S[g]$, the
second functional derivative with respect to the metric is different
from that with respect to the vierbein, see
\cite{Harst:2012ni,Harst:2013} for explicit representations on the
Einstein-Hilbert level. A second difference is more subtle:
quantizing the vierbein requires further gauge fixing of the
additional Lorentz symmetry. This gauge fixing goes along with
additional Faddeev-Popov ghosts. Though they can be ignored in
perturbation theory in the Lorentz symmetric gauge
\cite{Woodard:1984sj} as they are nonpropagating, they have been
shown to contribute nonperturbatively in
\cite{Harst:2012ni,Harst:2013}. In our spin-base invariant
formalism, there is no such artificial Lorentz symmetry and no
corresponding ghosts. Instead we have a local spin-base
invariance. As we have shown in the preceding section, the integral
over spin-bases factorizes in the functional integral in our
metric-based quantization scheme such that observables can be
computed in any desired spin base. Hence, we can just single out one
spin-base for the computation, e.g., by demanding
Eqs. \eqref{eq:optimal_choice_G_1} and \eqref{eq:optimal_choice_G_2}
to hold. Further ghosts could only appear if one wants to
explicitly carry out the integral over spin bases with (symbolic)
measure $\mathcal{D M}$ with a suitable spin-base gauge fixing. This
is, however, simply not necessary in the present formalism. From
another viewpoint, the choice of the spin basis as in
Eqs. \eqref{eq:optimal_choice_G_1} and \eqref{eq:optimal_choice_G_2}
plays the role of an external background field in our formalism
rather than a ``gauge''-fixed quantum field. Of course, other
choices are equally legimate as we have proved that spin-base
invariant observables do not depend on this choice.

\section{Conclusion and Outlook}
\label{sec:conc}

In this paper we gave a first-principles approach to a local spin-base
invariant approach to fermions in $4$ dimensional curved
spacetimes. While such a formalism already has been discussed
and successfully used at several instances in the literature, our
presentation carefully distinguishes between assumptions and
consequences, paving the way to generalizations and possibly
quantization.

One such generalization is the inclusion of torsion which we have
worked out for the first time in this article. In addition to
spacetime torsion, which can be included rather straightforwardly in
the formalism, the spin connection admits further degrees of freedom
which we interpret as spin torsion. Some of these degrees of freedom
can be associated with a scalar, an axial vector, and an
anti-symmetric tensor field. For instance, the latter has a coupling
to Dirac spinors in the form of a Pauli term. If the spin torsion
contains such a contribution, its torque-like physical influence on
the orientation of spin along a geodesic is
obvious. Phenomenologically, such terms are similar to those
discussed in standard model extensions due to Lorentz- or
CPT-violation \cite{Colladay:1998fq} and are typically tightly
constrained, see, e.g., \cite{Kostelecky:2008ts}.

Further generalizations include the construction of spin curvature
which can be used to define classical field theories of gravity (and
fermions) in terms of the Dirac matrices (and Dirac spinors) as
elementary degrees of freedom. We showed that the simplest possible
field theory contains Einstein's theory of general relativity and
predicts zero spacetime torsion and zero spin-torsion in absence of
explicit sources or boundary conditions.

For vanishing spacetime and spin torsion, the spin-base invariant
formalism can be mapped onto the conventional vierbein formalism which
can be viewed as ``spin-base gauge-fixed'' version of the invariant
formalism.

As another generalization, the formalism suggests the definition of a
generalized Lie derivative, which turns out to agree with the
generalized Lie derivative proposed by Kosmann. In our formalism, this
spinorial Lie derivative appears in a manner which can be given a
geometrical meaning much in the same way as the Lie derivative for
spacetime vectors can be associated with a geometrical interpretation.

As a main result, we used the formalism to show that a possible path
integral quantization of gravity and fermionic matter fields can be
solely based on an integration over metric and matter
fluctuations. Despite the fact that the Dirac matrices appear to be
the more fundamental degrees of freedom, their fluctuations can be
parametrized by metric as well as spin-base fluctuations. We observe
that the latter does not contribute to spin-base invariant observables
and hence the spin-base fluctuations can be factored out of the
quantum theory. In view of the increasing complexity of quantization
schemes based on vierbeins and/or spin connections, the legitimation
of a metric-based scheme (though still an open and frighteningly hard
challenge) is good news.

\acknowledgments 

The authors thank Martin Ammon, Felix Finster, Ulrich Harst, Ren\'{e}
Sondenheimer, Andreas Wipf, and Luca Zambelli for valuable
discussions, and Martin Ammon and Luca Zambelli for comments on the
manuscript. We acknowledge support by the DFG under grants Gi~328/5-2
(Heisenberg program), GRK1523 and FOR 723.


\appendix

\section{Spin metric}\label{App:spin_metric}

For a given set of Dirac matrices encoding the spacetime metric via
the Clifford-algebra constraint, also the spin metric $h$ is fixed (up
to a sign) by the requirements
\begin{align}
\begin{aligned}
 \rmi {}& \quad \gamma_{\mu}^{\dagger} = - h \gamma_{\mu} h^{-1} \text{,}\\
 \rmii {}& \quad \abs{\det h} = 1 \text{,}\\
 \rmiii {}& \quad h^{\dagger} = - h \text{.}
\end{aligned}
\end{align}
Let us first assume that there is at least one spin metric $h_{1}$,
which satisfies all three conditions. Then we know, if there is
another spin metric $h_{2}$, they must be related via
\begin{align}
 [ h_{2}^{-1} h_{1} , \gamma_{\mu} ] = 0 \text{,}
\end{align}
because both spin metrics have to fulfill
\begin{align}
 h_{2} \gamma_{\mu} h_{2}^{-1} = - \gamma_{\mu}^{\dagger} = h_{1} \gamma_{\mu} h_{1}^{-1} \text{.}
\end{align}
Therefore, using Schur's Lemma \cite{Schmutzer:1968},
\begin{align}
 h_{2} = z h_{1} , \quad z \in \C
\end{align}
has to hold. With $\rmii$, it follows that
\begin{align}
 z = \euler^{\cplx \arg z} \text{.}
\end{align}
But if both spin metrics satisfy the condition $\rmiii$, then
\begin{align}
 \euler^{- \cplx \arg z} h_{1} = - \euler^{-\cplx \arg z} h_{1}^{\dagger}  = - h_{2}^{\dagger} = h_{2} = \euler^{\cplx \arg z} h_{1}
\end{align}
has to hold. Therefore both spin metrics have to be identical up to a sign,
\begin{align}
 h_{2} = \pm h_{1} \text{.}
\end{align}
This demonstrates the uniqueness (up to a sign) of the spin
metric. Now we only need to prove the existence of one such spin metric
$h$. For this, we first introduce the Matrix $\hat{M}$ satisfying
\begin{align}
 \gamma_{\mu}^{\dagger} = - \euler^{\hat{M}} \gamma_{\mu} \euler^{- \hat{M}} , \quad \tr \hat{M} = 0 \label{eq:def_M} \text{.}
\end{align}
This equation implies
\begin{align}
 \gamma_{\mu}^{\dagger} = \euler^{\hat{M}} \gamma_{\ast} \gamma_{\mu} ( \euler^{\hat{M}} \gamma_{\ast} )^{-1}.
\end{align}
The matrices $\gamma_{\mu}^{\dagger}$ also satisfy the Clifford
algebra. Since any two different sets of such matrices satisfying the
Clifford algebra are connected by a similarity transformation
\cite{Schmutzer:1968}, i.e. a spin base transformation,
$\euler^{\hat{M}} \gamma_{\ast}$ must exist as it parametrizes this
similarity transformation. Therefore also $\hat M$ must exist but may
not be unique.  The trace of $\hat{M}$ can always be set to zero,
because the trace part commutes with all matrices and therefore drops
out of \Eqref{eq:def_M}.  The hermitean conjugate of \Eqref{eq:def_M}
is
\begin{align}
 \gamma_{\mu} = - \euler^{- \hat{M}^{\dagger}} \gamma_{\mu}^{\dagger} \euler^{\hat{M}^{\dagger}}.
\end{align}
Therefore, also 
\begin{align}
 \euler^{\hat{M}} \gamma_{\mu} \euler^{-\hat{M}} = - \gamma_{\mu}^{\dagger} = \euler^{\hat{M}^{\dagger}} \gamma_{\mu} \euler^{- \hat{M}^{\dagger}}
\end{align}
has to hold. Schur's Lemma again implies there exists a $\varphi$ such that
\begin{align}
 \euler^{\hat{M}^{\dagger}} = \euler^{\cplx \varphi} \euler^{\hat{M}} , \quad \varphi \in \R \text{.}
\end{align}
This equation fixes $\euler^{\cplx \varphi}$ once we have chosen a
specific $\hat{M}$. Now we also know, that $\det \euler^{\hat{M}} = 1$
and therefore the same has to hold for $\det
\euler^{\hat{M}^{\dagger}} = 1$. From this, we conclude that $\varphi$ is
limited to
\begin{align}
 \varphi \in \left\{ n \frac{2 \pi}{d_{\gamma}} : n \in \{ 0 , \ldots , d_{\gamma} - 1 \} \right\} \text{.}
\end{align}
The desired spin metric $h$ is then given by
\begin{align}
 h = \cplx \euler^{\cplx \frac{\varphi}{2}} \euler^{\hat{M}} \text{.}
\end{align}
It is straightforward to show, that this metric satisfies $\rmi$ - $\rmiii$.

We continue with implementing the spin metric compatibility
as expressed in \Eqref{eq:cov_deriv_spin_metric}. This tells us that
\begin{align}
 \Gamma_{\mu} + \bar{\Gamma}_{\mu} = h^{-1} \partial_{\mu} h
\end{align}
has to hold. Taking into account that (cf. \Eqref{eq:def_hatGamma})
\begin{align}
 - {D_\ttm{(\mathrm{LC})}}_{\mu} h \gamma^{\nu} h^{-1} \! = \! {D_\ttm{(\mathrm{LC})}}_{\mu} {\gamma^{\nu}}^{\dagger} \! = \! ({D_\ttm{(\mathrm{LC})}}_{\mu} \gamma^{\nu})^{\dagger} \! = - [ \hat{\Gamma}_{\mu} , \gamma^{\nu} ]^{\dagger},
\end{align}
we arrive at
\begin{align}
 [ h^{-1} ( \partial_{\mu} h ) - \hat{\Gamma}_{\mu} - \bar{\hat{\Gamma}}_{\mu} , \gamma^{\nu} ] = 0 \text{.}
\end{align}
Because $\tr \hat{\Gamma}_{\mu} = 0$, this implies
\begin{align}
 \hat{\Gamma}_{\mu} + \bar{\hat{\Gamma}}_{\mu} = h^{-1} \partial_{\mu} h - \frac{1}{d_{\gamma}} \tr ( h^{-1} \partial_{\mu} h ) \, \mrI \text{.}
\end{align}
Now we use
\begin{align}
 \tr \big( \euler^{- \hat{M}} \partial_{\mu} \euler^{\hat{M}} \big) ={}& \tr \! \left( \euler^{- \hat{M}} \sum\limits_{n = 1}^{\infty} \sum\limits_{k = 0}^{n - 1} \frac{\hat{M}^{k} (\partial_{\mu} \hat{M}) \hat{M}^{n - k - 1}}{n !} \right) \notag\\
 ={}& \tr ( \partial_{\mu} \hat{M} ) = 0
\end{align}
to conclude
\begin{align}
 \frac{1}{d_{\gamma}} \tr( h^{-1} \partial_{\mu} h) = \frac{\cplx}{2} \partial_{\mu} \varphi \text{.}
\end{align}
This leaves us with
\begin{align}
 \Gamma_{\mu} + \bar{\Gamma}_{\mu} = h^{-1} \partial_{\mu} h = \hat{\Gamma}_{\mu} + \bar{\hat{\Gamma}}_{\mu} + \frac{\cplx}{2} \partial_{\mu} \varphi \mrI \text{,}
\end{align}
which implies that
\begin{align}
 \frac{\cplx}{2} \partial_{\mu} \varphi = \frac{1}{d_{\gamma}} \tr ( \Gamma_{\mu} + \bar{\Gamma}_{\mu} ) = \frac{2}{d_{\gamma}} \Re \tr \Gamma_{\mu} \text{.}
\end{align}
Since the left-hand side is purely imaginary and the right-hand side
is purely real both have to vanish.  Because $\varphi$ can only take
discrete values, it must be a constant if we require it to be a
sufficiently smooth function. This finally implies that
\begin{align}
 \Re \tr \Gamma_{\mu} = 0
\end{align}
and
\begin{align}
 \Gamma_{\mu} + \bar{\Gamma}_{\mu} = \hat{\Gamma}_{\mu} + \bar{\hat{\Gamma}}_{\mu} = h^{-1} \partial_{\mu} h
\end{align}
have to hold. These two identities are used in Sect.~\ref{sec:Consequ}
to constrain spin torsion.

\section{Toolbox for the spin-base invariant formalism}\label{App:formulae}

In this appendix, we summarize a set of commonly used formulas for the
spin-base invariant formalism, which may serve as a toolbox for
practical computations. For simplicity, we set spacetime torsion and
spin torsion $\Delta \Gamma_\mu$ to zero.

Given a set of spacetime dependent Dirac matrices, the metric is
encoded in the Clifford-algebra constraint and can straightforwardly
be computed:
\begin{equation}
 \{ \gamma^{\mu} , \gamma^{\nu} \} = 2 \metric^{\mu \nu} \mrI , \quad
 g_{\mu \nu} = \frac{1}{d_{\gamma}} \tr(\gamma_{\mu} \gamma_{\nu}) .
\label{def:Clifford5}
\end{equation}
The inclusion of fermion degrees of freedom requires a spin metric $h$
for the definition of scalar products of spinors and conjugate spinors
\begin{align}
 \bar{\psi} = \psi^{\dagger} h \text{.}
\end{align}
Though $h$ can in principle be constructed explicitly,
cf. App. \ref{App:spin_metric}, only the algebraic relations that
define $h$ are typically needed in practical calculations,
\begin{align}
 {\gamma^{\mu}}^{\dagger} = - h \gamma^{\mu} h^{-1} \text{,} \quad \abs{\det h} = 1 \text{,} \quad h^{\dagger} = - h \text{.}
\end{align}
For covariant differentiation of spinors
\begin{align}
 \nabla_{\mu} \psi = \partial_{\mu} \psi + \Gamma_{\mu} \psi \text{,}
\end{align}
the affine spin connection is needed, where in the absence of spin
torsion ($\Delta \Gamma_\mu=0$) $\Gamma_{\mu}=\hat\Gamma_\mu$ is
implicitly given by
\begin{align}
 D_{\mu} \gamma^{\nu} = - [ \Gamma_{\mu} , \gamma^{\nu}] \text{,} \quad \tr \Gamma_{\mu} = 0,
\end{align}
and explicitly by
\begin{align}
\begin{aligned}
 \rmi \quad{}& \Gamma_{\mu} = p_{\mu} \gamma_{\ast} + v_{\mu}^{\point \alpha} \gamma_{\alpha} + a_{\mu}^{\point \alpha} \gamma_{\ast} \gamma_{\alpha} + t_{\mu}^{\point \alpha \beta} [ \gamma_{\alpha} , \gamma_{\beta} ] \text{,}\\
 \rmii \quad{}& p_{\mu} = \frac{1}{32} \tr(\gamma_{\ast} \gamma_{\alpha} \partial_{\mu} \gamma^{\alpha}) \text{,}\\
 \rmiii \quad{}& v_{\mu}^{\point \alpha} = \frac{1}{48} \tr( [ \gamma^{\alpha} , \gamma_{\beta} ] \partial_{\mu} \gamma^{\beta}) \text{,}\\
 \rmiv \quad{}& a_{\mu}^{\point \alpha} = \frac{1}{8} \tr( \gamma_{\ast} \partial_{\mu} \gamma^{\alpha} ) \text{,}\\
 \rmv \quad{}& t_{\mu \alpha}^{\point[2] \beta} = - \frac{1}{32} \tr( \gamma_{\alpha} \partial_{\mu} \gamma^{\beta} ) - \frac{1}{8} \christoffel{\beta \\ \mu \alpha} \equiv - t_{\mu \point \alpha}^{\point \beta} \text{.}
\end{aligned}
\end{align}
The covariant derivative satisfies the spin metric compatibility condition.
\begin{align}
 \nabla_{\mu} h = \partial_{\mu} h - h \Gamma_{\mu} - \Gamma_{\mu}^{\dagger} h = 0 \text{.}
\end{align}
The generalized Lie derivative $\tilde{\mcL}$ is given by
\begin{align}
 \tilde{\mcL}_{v} = \mcL_{v} \psi + \mcZ_{v} \psi\text{,}
\end{align}
where $\mcL_{v}$ is the ordinary Lie derivative acting on $\psi$ as on
a spacetime scalar, and the matrix $\mcZ_{v}$ is implicitly given by
\begin{align}
 \tilde{\mcL}_{v} \gamma^{\mu} = \mcL_{v} \gamma^{\mu} + [ \mcZ_{v} , \gamma^{\mu} ] = \frac{1}{2} (\mcL_{v} \metric^{\mu \nu}) \gamma_{\nu} \text{,} \quad \tr \mcZ_{v} = 0
\end{align}
and explicitly by
\begin{align}
 \mcZ_{v} = v^{\rho} \Gamma_{\rho} + \frac{1}{16} (\partial_{\rho} v_{\lambda} - \partial_{\lambda} v_{\rho}) [\gamma^{\rho} , \gamma^{\lambda}] \text{.}
\end{align}
For calculations in a quantized framework, the variations of the
spinorial quantities with respect to metric fluctuations $\delta
\metric_{\mu \nu}$ about a background metric $\bar{\metric}$ are
needed. Choosing a suitable spin base, these variations acquire a
minimal form (corresponding to the Lorentz symmetric gauge in the
vierbein formalism). Up to second order, the minimal variations are
given by
\begin{align}
 \metric_{\mu \nu} ={}& \bar{\metric}_{\mu \nu} + \delta \metric_{\mu \nu}\\
 \gamma_{\mu} ={}& \bar{\gamma}_{\mu} + \frac{1}{2} \delta \metric_{\mu \nu} \bar{\gamma}^{\nu} - \frac{1}{8} \delta \metric_{\mu \rho} \bar{\metric}^{\rho \lambda} \delta \metric_{\lambda \nu} \bar{\gamma}^{\nu} + \mcO(\delta \metric^3)\\
 h ={}& \bar{h} + \mcO(\delta \metric^3)\\
 \gamma_{\ast} ={}& \bar{\gamma}_{\ast} + \mcO(\delta \metric^3)\\
 \Gamma_{\mu} ={}& \bar{\Gamma}_{\mu} + \frac{1}{8} [ \bar{\gamma}^{\kappa} , \bar{\gamma}^{\sigma} ] \bar{D}_{\sigma} \delta \metric_{\kappa \mu} + \frac{1}{8} [ \bar{\gamma}^{\kappa} , \bar{\gamma}^{\sigma} ] \delta \metric_{\sigma \rho} \bar{\metric}^{\rho \lambda} \notag \\
 {}& \times \left( \frac{1}{4} \delta^{\nu}_{\mu} \delta^{\alpha \beta}_{\kappa \lambda} + \delta^{\alpha}_{\mu} \delta^{\nu}_{[\kappa} \delta^{\beta}_{\lambda]} \right) \! \bar{D}_{\nu} \delta \metric_{\alpha \beta} + \mcO(\delta \metric^3) \text{,}
\end{align}
where barred quantities refer to the background. 

The derivations of the identities of this toolboox as well as
generalizations to nonzero torsion can be found in the main text.


\begin{thebibliography}{89}

\bibitem{Ashtekar:2004eh} 
  A.~Ashtekar and J.~Lewandowski,
  Class.\ Quant.\ Grav.\  {\bf 21}, R53 (2004)
  [gr-qc/0404018]; 
%
  J.~F.~Plebanski,
  J.\ Math.\ Phys.\  {\bf 18}, 2511 (1977); 
%
  R.~Capovilla, T.~Jacobson and J.~Dell,
  Phys.\ Rev.\ Lett.\  {\bf 63}, 2325 (1989); 
  Class.\ Quant.\ Grav.\  {\bf 8}, 59 (1991); 
%
  K.~Krasnov,
  Phys.\ Rev.\ D {\bf 84}, 024034 (2011)
  [arXiv:1101.4788 [hep-th]]; 
  Phys.\ Rev.\ Lett.\  {\bf 106}, 251103 (2011)
  [arXiv:1103.4498 [gr-qc]].


\bibitem{Weinberg:1980gg}
  S.~Weinberg,
{\it  In *Hawking, S.W., Israel, W.: General Relativity*, 790-831}.

\bibitem{Reuter:1996cp}
  M.~Reuter,
  Phys.\ Rev.\  D {\bf 57}, 971 (1998)
  [arXiv:hep-th/9605030]; 
  M.~Niedermaier and M.~Reuter,
  Living Rev.\ Rel.\  {\bf 9}, 5 (2006); 
  R.~Percacci,
  In *Oriti, D. (ed.): Approaches to quantum gravity* 111-128
  [arXiv:0709.3851 [hep-th]]; 
  M.~Reuter and F.~Saueressig,
  New J.\ Phys.\  {\bf 14}, 055022 (2012)
  [arXiv:1202.2274 [hep-th]].

\bibitem{Harst:2012ni} 
  U.~Harst and M.~Reuter,
  JHEP {\bf 1205}, 005 (2012)
  [arXiv:1203.2158 [hep-th]].

\bibitem{Dona:2012am} 
  P.~Dona and R.~Percacci,
  Phys.\ Rev.\ D {\bf 87}, 045002 (2013)
  [arXiv:1209.3649 [hep-th]].


\bibitem{Daum:2010qt} 
  J.~-E.~Daum and M.~Reuter,
  Phys.\ Lett.\ B {\bf 710}, 215 (2012)
  [arXiv:1012.4280 [hep-th]]; 
  Annals Phys.\  {\bf 334}, 351 (2013)
  [arXiv:1301.5135 [hep-th]].

\bibitem{Harst:2013}
  U.~Harst, Dissertation, Mainz U. (2013)

\bibitem{Hebecker:2003iw} 
  A.~Hebecker and C.~Wetterich,
  Phys.\ Lett.\ B {\bf 574}, 269 (2003)
  [hep-th/0307109]; 
  C.~Wetterich,
  Phys.\ Rev.\ D {\bf 70}, 105004 (2004)
  [hep-th/0307145].


\bibitem{Diakonov:2011im} 
  D.~Diakonov,
  arXiv:1109.0091 [hep-th]; 
  A.~A.~Vladimirov and D.~Diakonov,
  Phys.\ Rev.\ D {\bf 86}, 104019 (2012)
  [arXiv:1208.1254 [hep-th]].

\bibitem{Finster:1997gn} 
  F.~Finster,
  J.\ Math.\ Phys.\  {\bf 39}, 6276 (1998)
  [hep-th/9703083].

\bibitem{Weldon:2000fr} 
  H.~A.~Weldon,
  Phys.\ Rev.\ D {\bf 63}, 104010 (2001)
  [gr-qc/0009086].
 
\bibitem{Weyl:1929}
  H.~Weyl, 
  Z.\ Phys.\ {\bf 56}, 330 (1929).

\bibitem{Fock:1929}
  V.~Fock and D.~Ivanenko,
  Compt.\ Rend.\ Acad.\ Sci.\ Paris {\bf 188}, 1470, (1929).

\bibitem{DeWitt:1965jb} 
  B.~S.~DeWitt,
  ``Dynamical theory of groups and fields,''
  Gordon \& Breach, New York, 1965.

\bibitem{Buchbinder:1992rb} 
  I.~L.~Buchbinder, S.~D.~Odintsov and I.~L.~Shapiro,
  ``Effective action in quantum gravity,''
  Bristol, UK: IOP (1992).

\bibitem{Watanabe:2004nt} 
  T.~Watanabe and M.~J.~Hayashi,
  gr-qc/0409029.
 
\bibitem{Finster:1998ws} 
  F.~Finster, J.~Smoller and S.~-T.~Yau,
  Phys.\ Rev.\ D {\bf 59}, 104020 (1999)
  [gr-qc/9801079].
 
\bibitem{Casals:2012es} 
  M.~Casals, S.~R.~Dolan, B.~C.~Nolan, A.~C.~Ottewill and E.~Winstanley,
  Phys.\  Rev.\ D {\bf 87}, 064027 (2013)
  [arXiv:1207.7089 [gr-qc]].

\bibitem{Floreanini:1989hq} 
  R.~Floreanini and R.~Percacci,
  Class.\ Quant.\ Grav.\  {\bf 7}, 975 (1990).

\bibitem{Lippoldt:2012}
  S.~Lippoldt,
  master thesis, Jena (2012).

\bibitem{Kosmann:1966}
  Y.~Kosmann,
  Comptes Rendus Acad.\ Sc.\ Paris, s\'{e}rie A {\bf 262}, 289 (1966);
  Y.~Kosmann,
  Comptes Rendus Acad.\ Sc.\ Paris, s\'{e}rie A {\bf 262}, 394 (1966);
  Y.~Kosmann,
  Comptes Rendus Acad.\ Sc.\ Paris, s\'{e}rie A {\bf 264}, 355 (1967);
  Y.~Kosmann,
  Ann.\ Mat. pura appl.\, IV, {\bf 91}, 317 (1972).

\bibitem{Fatibene:1996tf}  
  L.~Fatibene, M.~Ferraris, M.~Francaviglia and M.~Godina,
  gr-qc/9608003; 
  M.~Godina and P.~Matteucci,
  Int.\ J.\ Geom.\ Meth.\ Mod.\ Phys.\  {\bf 2}, 159 (2005).

\bibitem{Sharipov:2008xv} 
  R.~Sharipov,
  arXiv:0801.0622 [math.DG].

\bibitem{Schmutzer:1968}
  E.~Schmutzer,
  ``Relativistische Physik,''
  Leipzig, DE: Teubner Verlagsgesellschaft (1968).

\bibitem{Gies:2013dca}
  H.~Gies and S.~Lippoldt,
  Phys.\ Rev.\ D {\bf 87}, 104026 (2013)
  [arXiv:1303.4253 [hep-th]].

\bibitem{Woodard:1984sj} 
  R.~P.~Woodard,
  Phys.\ Lett.\ B {\bf 148}, 440 (1984).

\bibitem{vanNieuwenhuizen:1981uf} 
  P.~van Nieuwenhuizen,
  Phys.\ Rev.\ D {\bf 24}, 3315 (1981).
\bibitem{Eichhorn:2011pc} 
  A.~Eichhorn and H.~Gies,
  New J.\ Phys.\  {\bf 13}, 125012 (2011)
  [arXiv:1104.5366 [hep-th]].
\bibitem{Colladay:1998fq} 
  D.~Colladay and V.~A.~Kostelecky,
  Phys.\ Rev.\ D {\bf 58}, 116002 (1998)
  [hep-ph/9809521].
\bibitem{Kostelecky:2008ts} 
  V.~A.~Kostelecky and N.~Russell,
  Rev.\ Mod.\ Phys.\  {\bf 83}, 11 (2011)
  [arXiv:0801.0287 [hep-ph]].

\end{thebibliography}
\end{document}